\documentclass[preprintnumbers,article,amsmath,amssymb,floatfix,10pt,prd,onecolumn,
superscriptaddress,nofootinbib]{revtex4-2}
\usepackage{bm}
\usepackage{amsfonts}
\usepackage{latexsym}
\usepackage[latin1]{inputenc}
\usepackage{graphicx}
\usepackage{amsmath}
\usepackage{palatino}
\usepackage{mathpazo}
\usepackage{textcomp}
\usepackage{float}
\usepackage{booktabs}
\usepackage{dcolumn}
\usepackage{ragged2e}
\usepackage{hyperref}
\hypersetup{colorlinks,citecolor=blue}
\hypersetup{colorlinks=true,linkcolor=red,filecolor=magenta,    urlcolor=blue}
\usepackage{amsmath}
\usepackage{xcolor}
\usepackage{orcidlink}
\usepackage{epsfig}
\usepackage[caption=false]{subfig}
\usepackage{commath}
\def\jnl@style{\it}
\def\aaref@jnl#1{{\jnl@style#1}}

\def\aaref@jnl#1{{\jnl@style#1}}

\def\aj{\aaref@jnl{AJ}}                   % Astronomical Journal
\def\apj{\aaref@jnl{ApJ}}                 % Astrophysical Journal
\def\apjl{\aaref@jnl{ApJ}}                % Astrophysical Journal, Letters
\def\apjs{\aaref@jnl{ApJS}}               % Astrophysical Journal, Supplement
\def\apss{\aaref@jnl{Ap\&SS}}             % Astrophysics and Space Science
\def\aap{\aaref@jnl{A\&A}}                % Astronomy and Astrophysics
\def\aapr{\aaref@jnl{A\&A~Rev.}}          % Astronomy and Astrophysics Reviews
\def\aaps{\aaref@jnl{A\&AS}}              % Astronomy and Astrophysics, Supplement
\def\mnras{\aaref@jnl{Mon.~Not.~Roy.~Astron.~Soc.}}             % Monthly Notices of the RAS
\def\prd{\aaref@jnl{Phys.~Rev.~D}}        % Physical Review D
\def\prc{\aaref@jnl{Phys.~Rev.~C}}  % Physical Review C
\def\prl{\aaref@jnl{Phys.~Rev.~Lett.}}    % Physical Review Letters
\def\qjras{\aaref@jnl{QJRAS}}             % Quarterly Journal of the RAS
\def\skytel{\aaref@jnl{S\&T}}             % Sky and Telescope
\def\ssr{\aaref@jnl{Space~Sci.~Rev.}}     % Space Science Reviews
\def\zap{\aaref@jnl{ZAp}}                 % Zeitschrift fuer Astrophysik
\def\nat{\aaref@jnl{Nature}}              % Nature
\def\aplett{\aaref@jnl{Astrophys.~Lett.}} % Astrophysics Letters
\def\apspr{\aaref@jnl{Astrophys.~Space~Phys.~Res.}} % Astrophysics Space Physics Research
\def\physrep{\aaref@jnl{Phys.~Rep.}}      % Physics Reports
\def\physscr{\aaref@jnl{Phys.~Scr}}       % Physica Scripta
\def\commat{\aaref@jnl{Comm.~Math.~Phys.}}              % Communications in Mathematical Physics
\def\science{\aaref@jnl{Science}}               % Science
\def\cqg{\aaref@jnl{Classical Quant.~Grav.}}            % Classical and Quantum Gravity
\def\jpcs{\aaref@jnl{JPCS}}                                     % Journal of Physics Conference Series
\def\ijmpd{\aaref@jnl{Int.~J.~Mod.~Phys.~D}}                    % International Journal of Modern Physics D
\def\grg{\aaref@jnl{Gen.~Relat.~Gravit.}}               % General Relativity and Gravitation
\def\rpp{\aaref@jnl{Rep.~Prog.~Phys.}}          % Reports on Progress in Physics
\def\npa{\aaref@jnl{Nucl.~Phys.~A}}        % Nuclear Physics A
\def\lrr{\aaref@jnl{Living Rev.~Rel.}}                   % Living reviews in relativity
\def\jcap{\aaref@jnl{J.~Cosmology Astropart.~Phys.}}    % Journal of cosmology and astroparticle physics
\def\rmp{\aaref@jnl{Rev.~Mod.~Phys.}}   %Reviews of modern physics
\def\epjc{\aaref@jnl{Eur.~Phys.~J.~C}}
\def\plb{\aaref@jnl{~Phy.~Lett.~B}}
\def\mpla{\aaref@jnl{Mod.~Phy.~Lett.~A}}
\def\arxiv{\aaref@jnl{arxiv.org}}

\allowdisplaybreaks[1]
\begin{document}
%\color{red}
\color{black}       %% For one column
\title{A study of anisotropic spheres in $f(Q)$ gravity with quintessence field}
%\end{document}
\author{Sanjay Mandal\orcidlink{0000-0003-2570-2335}}
\email{sanjaymandal960@gmail.com}
\affiliation{Department of Mathematics, Birla Institute of Technology and
Science-Pilani,\\ Hyderabad Campus, Hyderabad-500078, India.}

\author{G. Mustafa}
\email{gmustafa3828@gmail.com}
\affiliation{Department of Physics, Zhejiang Normal University, Jinhua, 321004, People's Republic of China.}

\author{Zinnat Hassan\orcidlink{0000-0002-6608-2075}}
\email{zinnathassan980@gmail.com}
\affiliation{Department of Mathematics, Birla Institute of Technology and
Science-Pilani,\\ Hyderabad Campus, Hyderabad-500078, India.}

\author{P.K. Sahoo\orcidlink{0000-0003-2130-8832}}
\email{pksahoo@hyderabad.bits-pilani.ac.in}
\affiliation{Department of Mathematics, Birla Institute of Technology and
Science-Pilani,\\ Hyderabad Campus, Hyderabad-500078, India.}
%
%%%%%%%%%%%%%%%%%%%%%%%%%%%%%%%%%%%%%  DATE  %%%%%%%%%%%%%%%%%%%%%%%%%%%%%%%%%%%%
\date{\today}
\begin{abstract}
This manuscript examines the compact stars in gravity $f (Q)$, where non-metricity $Q$ drives gravitational interactions. To achieve this goal, we consider a spherically symmetric spacetime with the anisotropic fluid distribution. In particular, the quintessence field is used in the energy-momentum tensor to explore the solution of compact stars. In addition, we select three specific types of compact stars, namely HerX-1, SAXJ1808.4-3658, and 4U1820-30. We develop the field equations using the quintessence field of a specific form of $f(Q)$ as $f(Q)=Q-\alpha \gamma \left(1-\frac{1}{e^{\frac{Q}{\gamma}}}\right)$  in $f (Q)$ gravity. In addition, we consider the Schwarzschild metric for the matching conditions at the boundary. Then, we calculate the values of all the relevant parameters by imposing matching conditions. We provide detailed graphical analyses to discuss the parameters' physical viability, namely energy density, pressure, gradient and anisotropy. We also examine the stability of compact stars by testing energy conditions, equations of state, conditions of causality, redshift functions, mass functions, and compactness functions. Finally, we find that the solutions we obtained are physically feasible in modified $f (Q)$ gravity, and also it has good properties for the compact star models.
\end{abstract}

\maketitle

\date{\today}
\section{Introduction}

The debate over compact stars has raged on since the accelerated expansion of the universe was discovered. Compact stars are quite different from ordinary stars, which are come into existence after the death of ordinary stars. All the nuclear processes that took place within the cores of ordinary stars have ended. The gravitational force inside the star is more dominant than the repulsive force because the temperature decreases inside the star. This phenomena ends with the compact stars formation, and all the masses of the stars accumulate in a very short radius. Therefore, compact stars are seen as high-density masses with a compact radius. Due to numerous problems in astrophysics, the discussion of compact stars has aroused great interest among cosmological researchers today. There are many reference points that can be used to study compact stars; however, there are some ambiguities because the perfect formation of compact stars remains problematic. For example, we can analyze the balance of compact stars by discussing the TOV equation proposed by Tollman-Oppenheimer-Volkoff. In fact, the balance characteristics of TOV forces, namely anisotropy, gravity, and hydrostatic force, indicate that the dense object system we are studying is stable and physically acceptable. Many compelling efforts by researchers can be found in the literature \cite{folomeev/2012}.

Despite the great success of Newtonian gravity, it completely failed under certain circumstances considering strong gravitational effects, such as the progress of Mercury other than the Mickelson Morey experiment \cite{Eisele/2009}. The General Theory of Relativity (GR) was developed by Einstein in 1915, which made it possible to solve the Mercury problem \cite{wheeler/1990}. Since then, GR is considered to be the basic theory of gravitational physics. However, GR has several shortcomings, which show that GR is not the complete fundamental theory of gravity, such as the dark energy problem \cite{perlmutter/1999}. Furthermore, GR violated the Chandrasekhar mass restriction of the Super Chandrasekhar white dwarf and the sub-Chandrasekhar quality restriction \cite{howell/2006}. Furthermore, GR shows inconsistencies in the strong gravitational field and recent observations \cite{Hawkins/2003}. Therefore, look for the appropriate modifications to the GRs. It is a proactive task. The most successful modification of GR is the theory of higher-order curvature, especially gravity $f (R)$, which successfully explains the existence of dark matter and uses observations to counter the theory of gravity \cite{Bamba/2012}. Besides this, teleparallel gravity and symmetric teleparallel gravity provide an interesting geometric interpretation of gravity with the spacetime metric. Recently, symmetric teleparallel gravity ($f(Q)$ gravity) was proposed by Jim\'enez et al \cite{Jemenez/2018}, which is a well-motivated theory of gravity, where the non-metricity drive the gravitational interaction in spacetime.

Moreover, the stellar structures study in modified theories of gravity is an important topic for discussion. In the last few decades, several studies presented on stellar structures. However, recently, the compact star models in $f(R)$ gravity was studied by Nashed and Capozziello \cite{nashed/2021}, there the stability of compact star tested using TOV(Tolman-Oppenheimer-Volkoff) equation, and their solution matches with the observational data of millisecond pulsars with a withe dwarf companion and pulsars presenting thermonuclear bursts. Astashenok and Odintsov \cite{Astashenok/2020} investigated the rotating newton stars with axion field in $f(R)$ gravity and discussed stars with different frequencies and masses. The researchers used an interesting method to derive the solution of the field equation in the context of compact objects, the Karmarkar condition. This condition was first proposed by Karmarkar \cite{Karmarker/1948}, and it is considered a mandatory condition for type I embedding in spherically symmetric space-time. It can help us to find the exact solution of the motion equations. Using the Karmarkar condition, several works have been done in modified gravity to study the stellar structures (see details in the references here \cite{Mustafa/2020a,Mustafa/2020b,Mustafa/2020}).  Furthermore, some interesting studies have been done in theories of gravity to explore the stellar structure solutions \cite{Kalam/2014}.

Since $ f (Q) $ gravity was proposed, the research on it has been increasing rapidly. At the same time, it is faced with the observational limitation of using GR's standard formula to counter it. Lazkoz et al. impose an compelling set of restrictions on gravity $f (Q)$. \cite{Lazkoz/2019}, where Lagrange $f (Q)$ is expressed as a polynomial function of redshift $z$. Using observational data such as Type Ia supernovae, baryon acoustic oscillation data, quasars, gamma-ray bursts, and cosmic microwave background distance data, the bounds of these models have been derived successfully. Moreover, some interesting works have been done in $f(Q)$ gravity using observational measures recently \cite{mandal/2020}. Besides this, $f(Q)$ gravity is also developed to study the wormhole solutions \cite{Hasan/2021}, its signature \cite{Frusciante/2021}, Spherically symmetric configuration \cite{Lin/2021}, and so on. As this theory was developed recently, it is quite successful in describing the current scenario of the universe. This success of $f(Q)$ gravity motivates us to test the compact star's stability and present physically acceptable models in the presence of quintessence filed in the fluid description. Therefore, exploring the stellar structures in the $f(Q)$ gravity background would be interesting as it is a novel approach in $f(Q)$ gravity. Here, we aim to explore the stellar structures in $f(Q)$ gravity with the presence quintessence field.

This manuscript is organized as follows: in Section \ref{sec2}, we derive the field equations in the background of $f(Q)$ gravity with the presence of quintessence field in the anisotropic fluid description. Also, there we considered the Krori and Barua gravitational potentials and Herrera and Barreto equation of state. In Section \ref{sec3}, we use the matching condition to find the values of parameters, which are used in our formulation. In Section \ref{sec4}, we discuss the physical analysis of our constructed star models such as HerX-1, SAXJ1808.4-3658, and 4U1820-30. The physical analysis such as energy density and pressure components under quintessence field in \ref{sec4a}, equation of state parameters in \ref{sec4b}, Gradients in \ref{sec4c}, anisotropy in \ref{sec4d}, causality and Abreu stability analysis in \ref{sec4e}, energy conditions in \ref{sec4f}, mass function, compactness function, and redshift function in \ref{sec4g} are discussed to present a physically viable compact star solutions. Finally, combining all the outcomes, we have concluded our study for the compact star in Section \ref{sec5}.

\section{Model of anisotropic quintessence star in $f(Q)$ gravity}\label{sec2}

We have considered the activity for symmetric teleparallel gravity as \cite{Jemenez/2018}
\begin{equation}\label{1}
\mathcal{S}=\int\left(\frac{1}{2}\,f(Q)\,+ \mathcal{L}_m\,\right)\sqrt{-g}\,d^4x\, ,
\end{equation}
where the matter Lagrangian density represented by $\mathcal{L}_m$, $g$ is the determinant of the metric $g_{\mu\nu}$, and $f(Q)$ addresses the arbitrary function of $Q$,.\\
The non-metricity tensor and its traces can be written as\\
\begin{equation}\label{2}
Q_{\lambda\mu\nu}=\bigtriangledown_{\lambda} g_{\mu\nu},
\end{equation}
\begin{equation}
\label{3}
 \tilde{Q}_\alpha=Q^\mu\;_{\alpha\mu},\, \, Q_{\alpha}=Q_{\alpha}\;^{\mu}\;_{\mu}.
\end{equation}
Also, the superpotential in terms of the non-metricity tensor can be written as
\begin{equation}\label{4}
P^\alpha\;_{\mu\nu}=\frac{1}{4}\left[-Q^\alpha\;_{\mu\nu}+2Q_{(\mu}\;^\alpha\;_{\nu)}+Q^\alpha g_{\mu\nu}-\tilde{Q}^\alpha g_{\mu\nu}-\delta^\alpha_{(\mu}Q_{\nu)}\right],
\end{equation}
here the trace of non-metricity tensor \cite{Jemenez/2018} takes the following expression
\begin{equation}
\label{5}
Q=-Q_{\alpha\mu\nu}\,P^{\alpha\mu\nu}.
\end{equation}
The fluid description of spacetime can be written in terms of energy-momentum tensor as
\begin{equation}\label{6}
T_{\mu\nu}=-\frac{2}{\sqrt{-g}}\frac{\delta\left(\sqrt{-g}\,\mathcal{L}_m\right)}{\delta g^{\mu\nu}}.
\end{equation}
Now, by shifting the action \eqref{1} with respect to the metric tensor $g_{\mu\nu}$ one can compose the motions equation, which can be written as
\begin{equation}\label{7}
\frac{2}{\sqrt{-g}}\bigtriangledown_\gamma\left(\sqrt{-g}\,f_Q\,P^\gamma\;_{\mu\nu}\right)+\frac{1}{2}g_{\mu\nu}f \\
+f_Q\left(P_{\mu\gamma i}\,Q_\nu\;^{\gamma i}-2\,Q_{\gamma i \mu}\,P^{\gamma i}\;_\nu\right)=-T_{\mu\nu},
\end{equation}

where $f_Q=\frac{df}{dQ}$.\\
Also, one can obtain the following relation by  varying \eqref{1} with respect to the connection,
\begin{equation}\label{8}
\bigtriangledown_\mu \bigtriangledown_\nu \left(\sqrt{-g}\,f_Q\,P^\gamma\;_{\mu\nu}\right)=0.
\end{equation}

The curvature less and torsionless make the affine connection to the following form
\begin{equation}
\label{50}
\Gamma^{\alpha}_{\mu\nu}=\left(\frac{\partial x^{\alpha}}{\partial\xi^{\lambda}}\right)\partial_\mu \partial_\nu \xi^{\lambda}
\end{equation}

Here we use a special coordinate choice, called coincident gauge so that the affine connection, $\Gamma^{\alpha}_{\mu\nu}=0$\\
and hence, the non-metricity reduces to
\begin{equation}
\label{51}
Q_{\alpha\mu\nu}=\partial_\alpha g_{\mu\nu},
\end{equation}
and thereby largely simplifies the calculation since only the metric is the fundamental variable.\\
Since the affine connection in Eq. \eqref{50} is purely inertial, one could utilize the covariant formulation by first determining affine connection in the absence of gravity \cite{Zhao}.\\
Here we built stellar structures by taking the spherically symmetric spacetime. This spacetime is conventionally composed as
\begin{equation}\label{9}
ds^2=e^{2\,a(r)}dt^2-e^{2\,b(r)}dr^2-r^2\left( d\theta^2+\text{sin}^2\theta d\phi^2\right),
\end{equation}
where $a(r)$ and $b(r)$ solely depend on the radial coordinate r. For the current interest, we consider the matter is depicted by an anisotropic stress-energy tensor of the structure
\begin{equation}\label{10}
T_{\mu}^{\nu}=\left(\rho+p_t\right)u_{\mu}\,u^{\nu}-p_t\,\delta_{\mu}^{\nu}+\left(p_r-p_t\right)v_{\mu}\,v^{\nu},
\end{equation}
where $\rho$, $P_r$ and $P_t$ correspond to energy density, radial and transverse pressures, respectively. The expressions $u^{\mu}$ and $v^{\nu}$ are the four-velocity vector and the unitary space-like vector in the radial direction, respectively and $u^{\mu}=e^a \delta_{0}^{\mu}$, $v^{\mu}=e^b \delta_{0}^{\mu}$. Now, we define the energy momentum tensor for quintessence field, which is realized as:
\begin{equation}\label{11}
\tilde{T}_{\mu}^{\nu}=T_{\mu}^{\nu}+{\mathcal{D}}_{\mu}^{\nu}
\end{equation}
where ${\mathcal{D}}_{\mu}^{\nu}$ is a energy momentum tensor for quintessence field. Further, we get the following relations from the energy momentum tensor for quintessence field.
\begin{equation}\label{12}
{\mathcal{D}}_{t}^{t}={\mathcal{D}}_{r}^{r}=-\rho_{q}
\end{equation}
and
\begin{equation}\label{13}
{\mathcal{D}}_{\theta}^{\theta}={\mathcal{D}}_{\phi}^{\phi}=\frac{(3 w_{q}+1)\rho_{q}}{2}
\end{equation}
where $\rho_{q}$ represents the energy density for quintessence filed $w_{q}$, with ($-1<w_{q}<-\frac{1}{3}$). For the metric \eqref{9},the trace of the non-metricity tensor $Q$ is given below,
\begin{equation}\label{14}
Q=\frac{2}{r}e^{-2\,b(r)}\left(2\,a^{'}(r)+\frac{1}{r}\right).
\end{equation}
Now, we define an exponential type function for $f(Q)$ gravity, which is expressed as:
\begin{equation}\label{15}
f(Q)=Q-\alpha \gamma \left(1-\frac{1}{e^{\frac{Q}{\gamma}}}\right)
\end{equation}
where $\alpha\in(0,1)$ and $\gamma=\Omega H^{2}_{0}$ with $\Omega$ is dimensionless parameter and $H^{2}_{0}$ is a Hubble parameter. By using the symmetry of the system we can choose the Krori and Barua gravitational potentials \cite{Mustafa/2020}, which are defined as:
\begin{equation} \label{16}
a(r)=Ar^2\;\;\;\;\;\;\;\;\;\;\;\;\;\;b(r)=Br^2+C
\end{equation}
where $A$, $B$, $C$ are the arbitrary constants. By using Eqs. \eqref{16} with quintessence field we get a following field equations
\begin{eqnarray}
\rho -\rho_q &=&\frac{e^{-2 A r^2} \left(r^2 +e^{A r^2} \left(e^{A r^2} \left(\alpha  \gamma  r^2+2\right)+4 A r^2-2\right)-\frac{\alpha  \chi_{1} \left(r^2 \chi _2+16\right)}{\gamma }\right)}{2 r^4}, \label{17}\\
p_r+\rho_q&=&\frac{e^{-A r^2} \left(\alpha  \chi _1 \left(e^{A r^2} \left(\gamma  r^2+2\right)-8 B r^2-4\right)-e^{A r^2} \left(\alpha  \gamma  r^2+2\right)+4 B r^2++2\right)}{2 r^2},\label{18}\\
p_t-\frac{1}{2}  (3w_q+1)\rho_q&=&\frac{-2 r^4 \chi _3 \left(\alpha  \chi _1-1\right) e^{-A r^2}+r^4 e^{-A r^2} \left(r^2 \chi _5+2\right)-\chi _4}{2 r^6}\label{19}.
\end{eqnarray}
where
\begin{eqnarray*}
\chi _1&&=e^{\frac{e^{-A r^2} \left(4 B r^2+2\right)}{\gamma  r^2}},\\
\chi _2&&=4 \gamma  e^{A r^2} \left(r^2 (A-B)-1\right)+16 A \left(2 B r^2+1\right)+\gamma  e^{2 A r^2} \left(\gamma  r^2+2\right),\\
\chi _3&&=B r^4 (A-B)+r^2 (A-4 B)-1,\\
\chi _4&&=\frac{8 \alpha  r^2 \left(B r^2+1\right) \left(A \left(2 B r^4+r^2\right)+1\right) \exp \left(\frac{e^{-A r^2} \left(4 B r^2+2\right)}{\gamma  r^2}-2 A r^2\right)}{\gamma },\\
\chi _5&&=4 B-\alpha  \gamma  \left(\chi _1-1\right) e^{A r^2}.
\end{eqnarray*}
To express the explicit relations between energy density and radial pressure, known as the equation of state. In the current analysis, the following linear equation of state is considered:
\begin{equation}\label{20}
p_r=\omega  \rho, \;\;\;\;\;\;\;\;\;\;\;0<\omega<1.
\end{equation}
where $\omega$ represents the equation of state parameter. Herrera and Barreto \cite{Herrera/2013} have introduced the general form of this relation.\\
By using Eq. (\ref{20}) in Eqs. (\ref{17}-\ref{19}), we get the following expressions:
\begin{eqnarray}
\rho&=&-\frac{2 (A+B) e^{-A r^2}-\frac{\exp \left(\frac{e^{-A r^2} \left(4 B r^2+2\right)}{\gamma  r^2}-2 A r^2\right) \left(2 \alpha  \gamma  r^4 (A+B) e^{A r^2}+8 \alpha  \left(A \left(2 B r^4+r^2\right)+1\right)\right)}{\gamma  r^4}}{-\omega -1}, \label{21}\\
p_r&=&-\frac{\omega  \left(2 (A+B) e^{-A r^2}-\frac{\exp \left(\frac{e^{-A r^2} \left(4 B r^2+2\right)}{\gamma  r^2}-2 A r^2\right) \left(2 \alpha  \gamma  r^4 (A+B) e^{A r^2}+8 \alpha  \left(A \left(2 B r^4+r^2\right)+1\right)\right)}{\gamma  r^4}\right)}{-\omega -1},\label{22}\\
\rho_q&=&-\frac{2 (A+B) e^{-A r^2}-\frac{\exp \left(\frac{e^{-A r^2} \left(4 B r^2+2\right)}{\gamma  r^2}-2 A r^2\right) \left(2 \alpha  \gamma  r^4 (A+B) e^{A r^2}+8 \alpha  \left(A \left(2 B r^4+r^2\right)+1\right)\right)}{\gamma  r^4}}{\omega +1}\nonumber\\&-&\frac{e^{-2 A r^2} \left(\frac{\alpha  \chi _1 \left(r^2 \chi _2+16\right)}{\gamma }-r^2 e^{A r^2} \left(e^{A r^2} \left(\alpha \gamma  r^2+2\right)+4 A r^2-2\right)\right)}{2 r^4},\label{23}\\
p_t&=&-\frac{-2 r^4 \chi _3 \left(\alpha  \chi _1-1\right) e^{-A r^2}+r^4 e^{-A r^2} \left(r^2 \chi _5+2\right)-\chi _4}{2 r^6}-\frac{1}{2} \text{$\rho $q} (3 w_{q}+1)\label{24}.
\end{eqnarray}

\section{Matching Conditions}\label{sec3}
In order, to find the values of involved unknowns, we shall match the interior spacetime with the exterior spacetime. For this purpose, we shall use the the Schwarzschild spacetime, which is defined as:
\begin{equation}\label{18}
ds^2=\left(1-\frac{2M}{r}\right)dt^2-\left(1-\frac{2M}{r}\right)^{-1}dr^2-r^2\left(d\theta^2+\sin^2\theta d\phi^2\right).
\end{equation}
where $M$ is termed as the total mass confined within the star and $R_{\varepsilon}$ to be the radius of the compact stellar remnant. The following expressions are carried at $r=R$ by considering the metric potentials:
\begin{eqnarray}
&&g_{tt}^-=g_{tt}^+,\;\;\;\;\;\;\;g_{rr}^-=g_{rr}^+,~~~~\frac{\partial g_{tt}^- }{\partial r}=\frac{\partial g_{tt}^+}{\partial r}.\label{26}
\end{eqnarray}
The signature of the intrinsic and extrinsic are illustrated as $(-)$ and $(+)$ respectively, i.e. $r=R_{\varepsilon}$. The values of the required constraints are retrieved by comparing the interior and exterior matric as are and termed out to be as following:
\begin{eqnarray}
&&A=-\frac{1}{R_{\varepsilon}^2}log(1-\frac{2M}{R_{\varepsilon}}),~~~~B=-\frac{M}{R_{\varepsilon}^2}log(1-\frac{2M}{R_{\varepsilon}})^{-1}\label{28},
~~~C=log(1-\frac{2M}{R_{\varepsilon}})-\frac{M}{R_{\varepsilon}}(1-\frac{2M}{R_{\varepsilon}})^{-1}.
\end{eqnarray}

\begin{table}[h]
\caption{\label{tab1}{Approximated values of involved parameters with $(\gamma =-0.02)$, $\omega =0.003;)$ and $w_{q}=-0.0.5;)$.}}
\vspace{0.6cm}\begin{tabular}{|c|c|c|c|c|c|}
\hline
$Strange Quark Star$ &\textbf{$M$} &\textbf{$R_{\varepsilon}(km)$} &\textbf{$\frac{M}{R_{\varepsilon}}$} &\textbf{$A(km^{-2})$} &\textbf{$B(km^{-2})$} \\
\hline
$Her X-1$&$0.88M_\odot$&$7.7$&$0.168$&$0.0069062764$&$0.0042673646$ \\
$SAX J 1808.4-3658$&$1.435M_\odot$&$7.07$&$0.299$&$0.018231569$&$0.014880115$ \\
$4U 1820-30$&$2.25M_\odot$&$10.0$&$0.332$&$0.010906441$&$0.0098809523$\\
\hline
\end{tabular}
\end{table}
\section{Physical Analysis}\label{sec4}
In this section, we shall discuss the necessary conditions with the required behavior for the stellar system.

\subsection{Energy density and pressure components under quintessence filed}\label{sec4a}
\begin{figure}[H]
\centering \epsfig{file=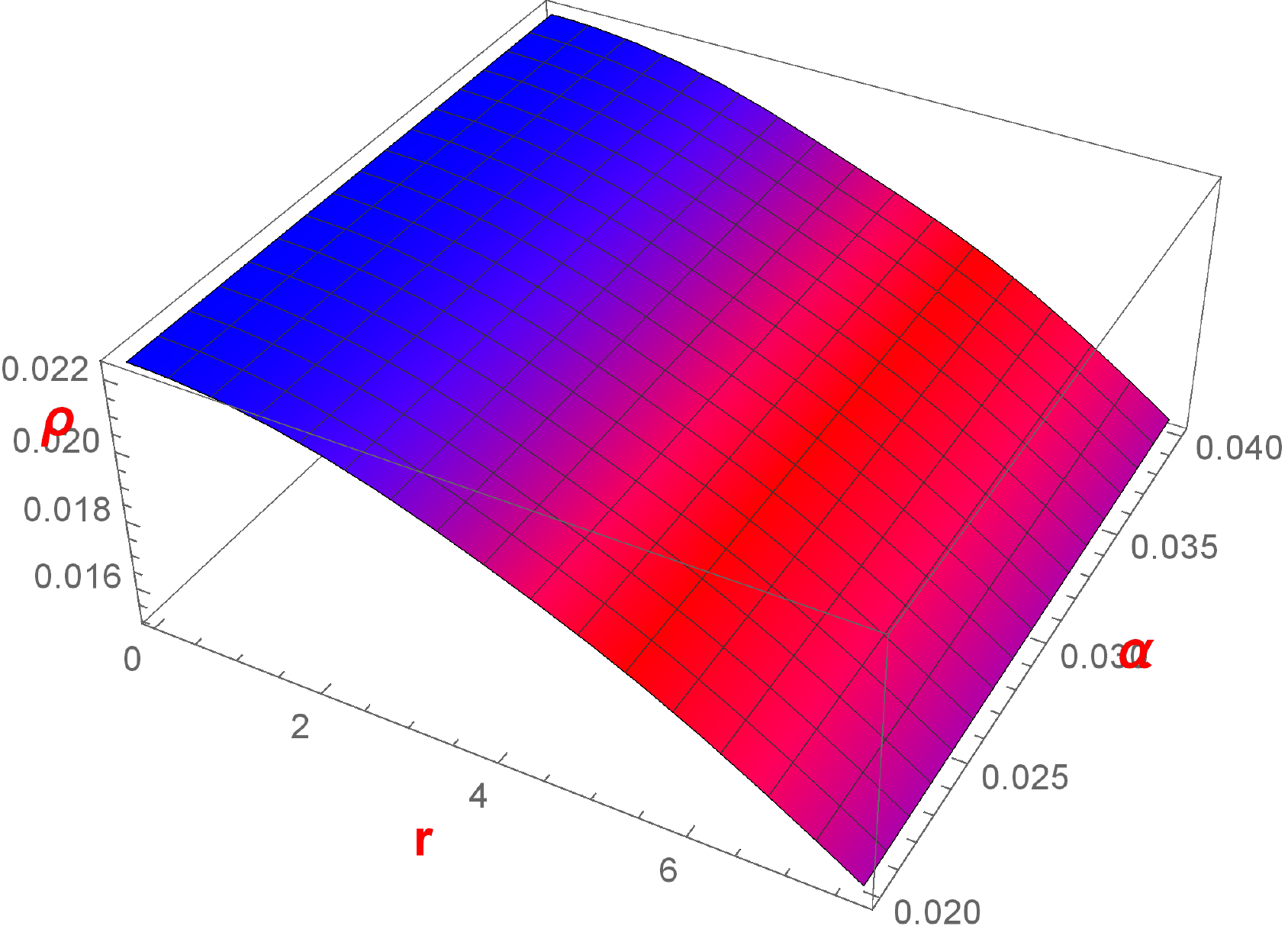, width=.32\linewidth,
height=2.5in}\epsfig{file=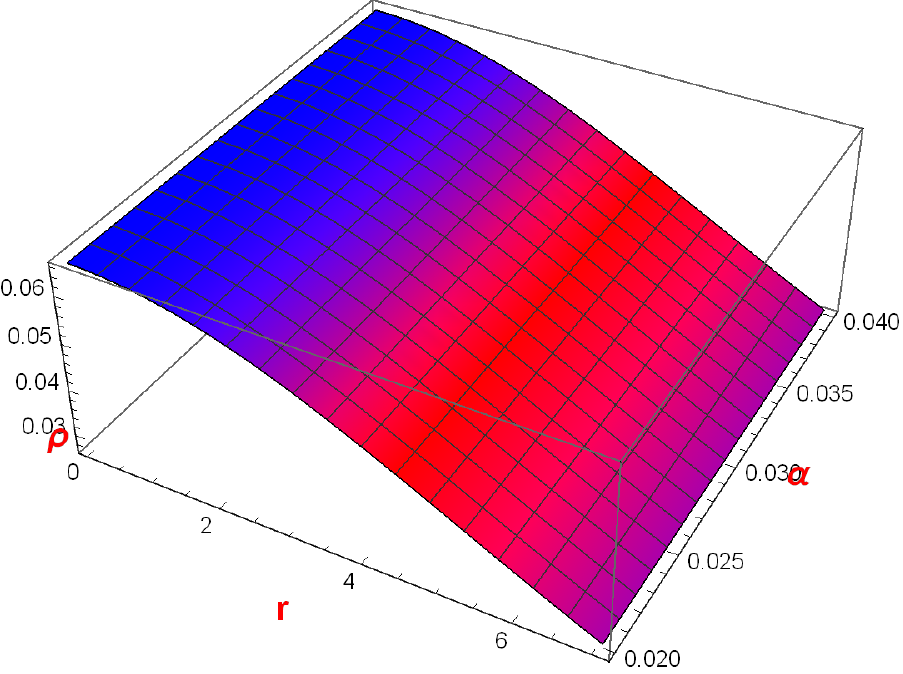, width=.32\linewidth,
height=2.5in}\epsfig{file=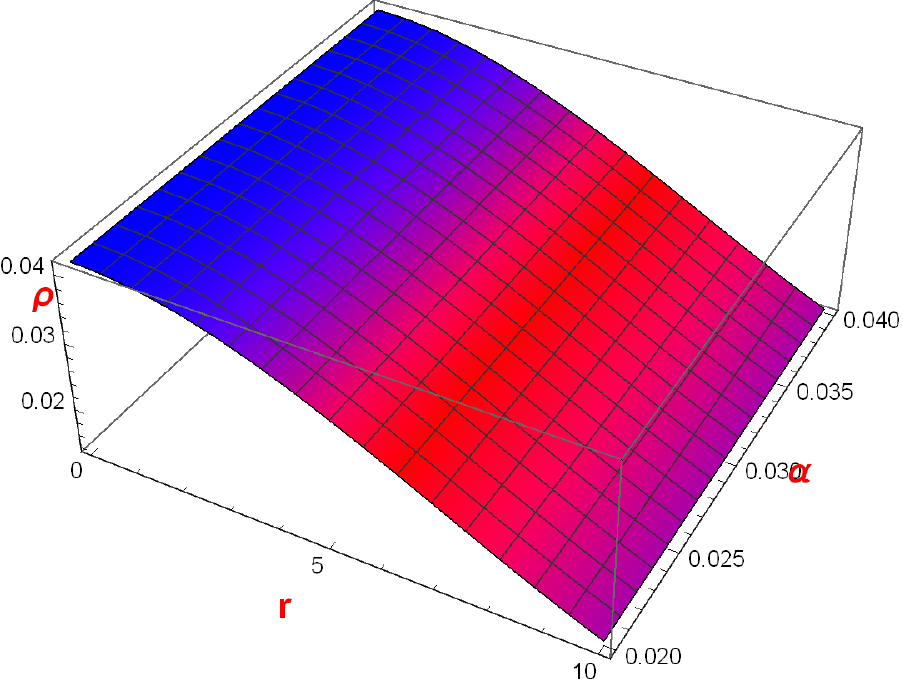, width=.32\linewidth,
height=2.5in}\caption{\label{Fig.1} shows the behavior of energy density.}
\end{figure}

\begin{figure}[H]
\centering \epsfig{file=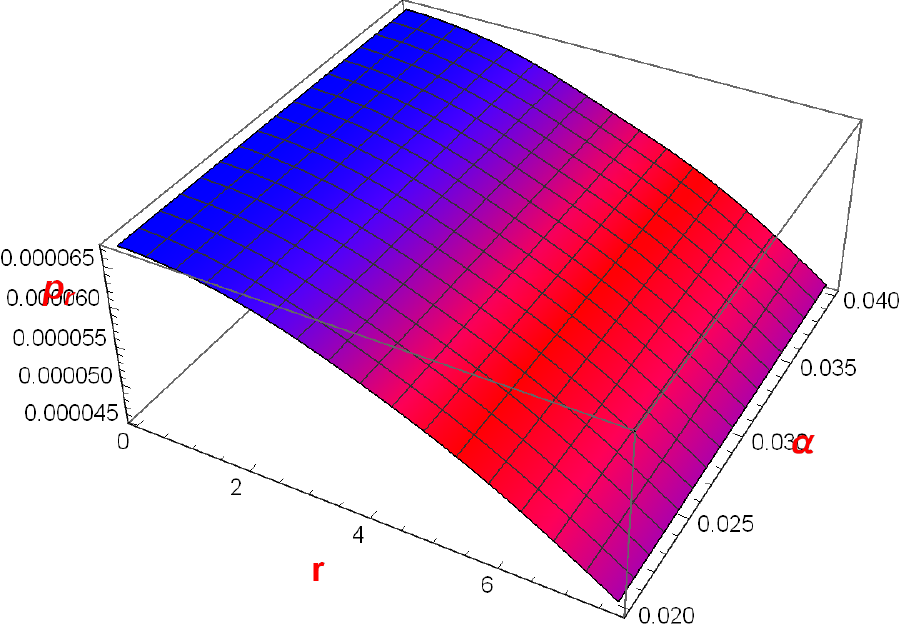, width=.32\linewidth,
height=2.5in}\epsfig{file=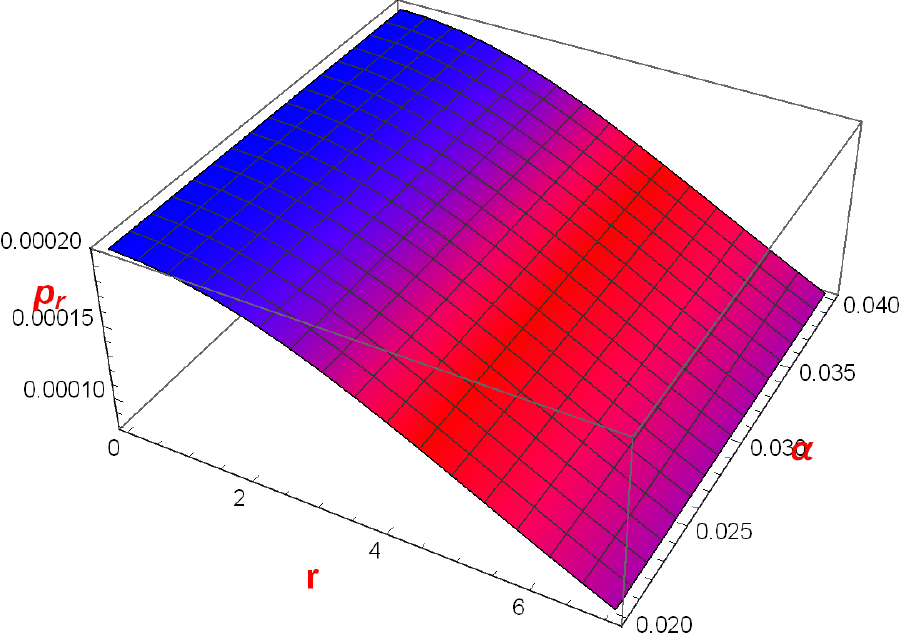, width=.32\linewidth,
height=2.5in}\epsfig{file=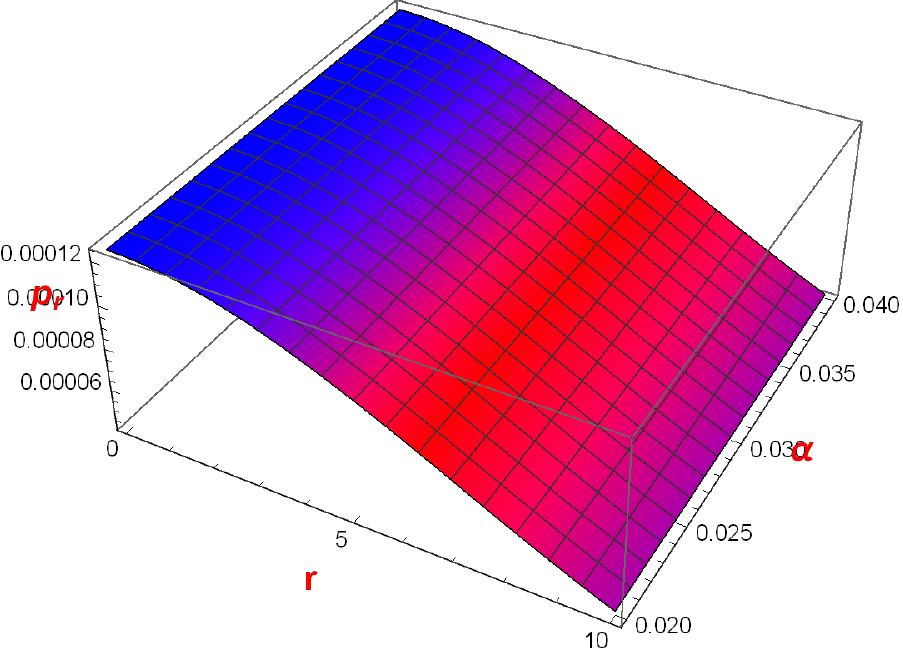, width=.32\linewidth,
height=2.5in}\caption{\label{Fig.2} shows the behavior of radial pressure.}
\end{figure}

\begin{figure}[H]
\centering \epsfig{file=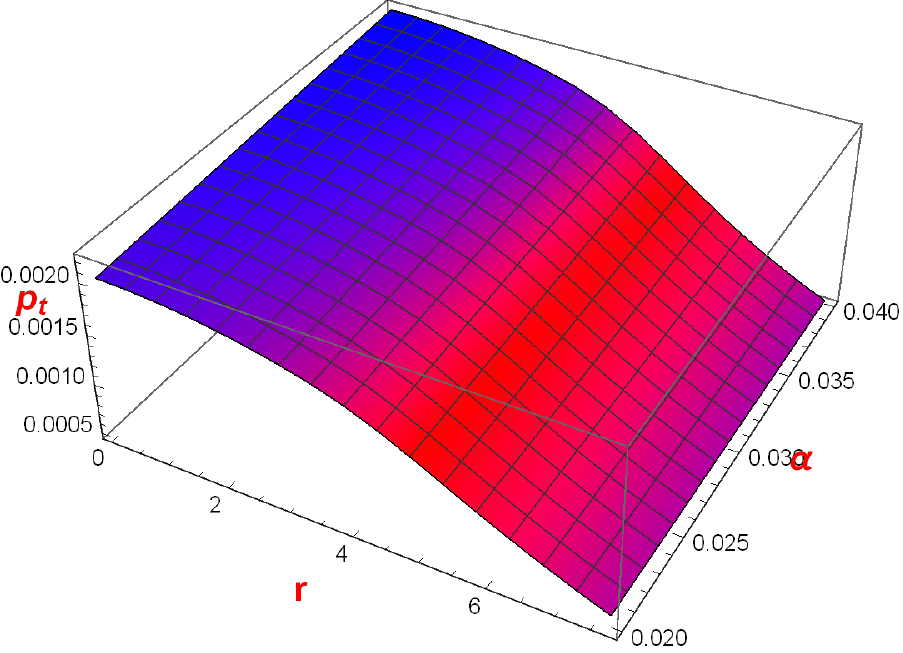, width=.32\linewidth,
height=2.5in}\epsfig{file=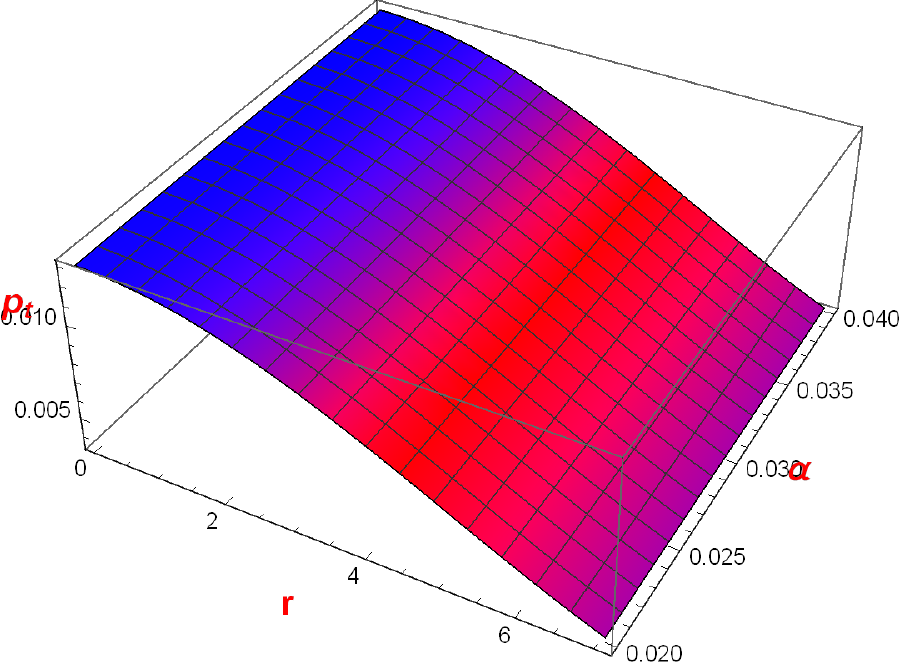, width=.32\linewidth,
height=2.5in}\epsfig{file=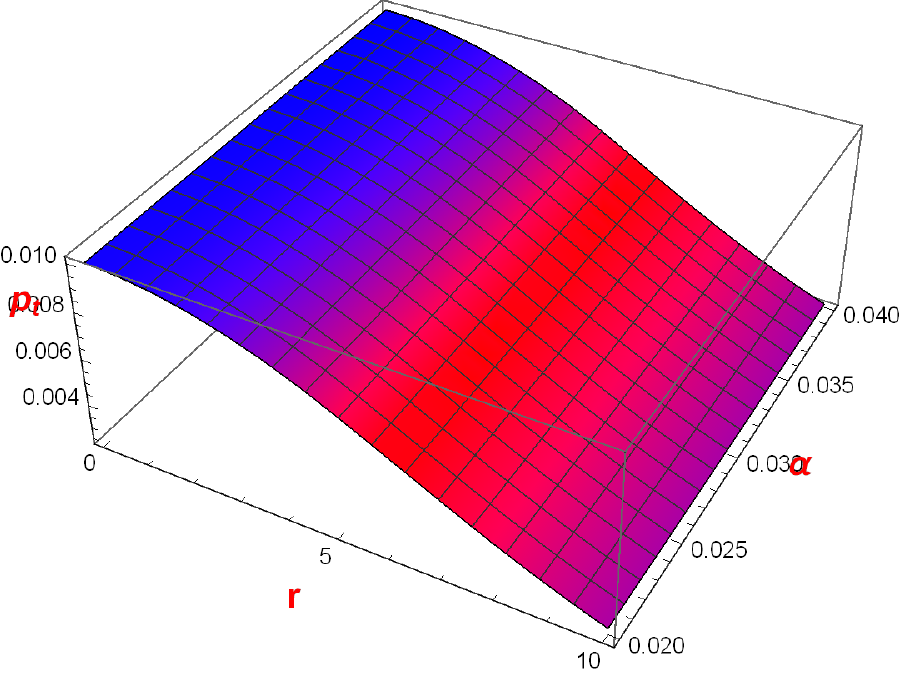, width=.32\linewidth,
height=2.5in}\caption{\label{Fig.3} shows the behavior of tangential pressure.}
\end{figure}

\begin{figure}[H]
\centering \epsfig{file=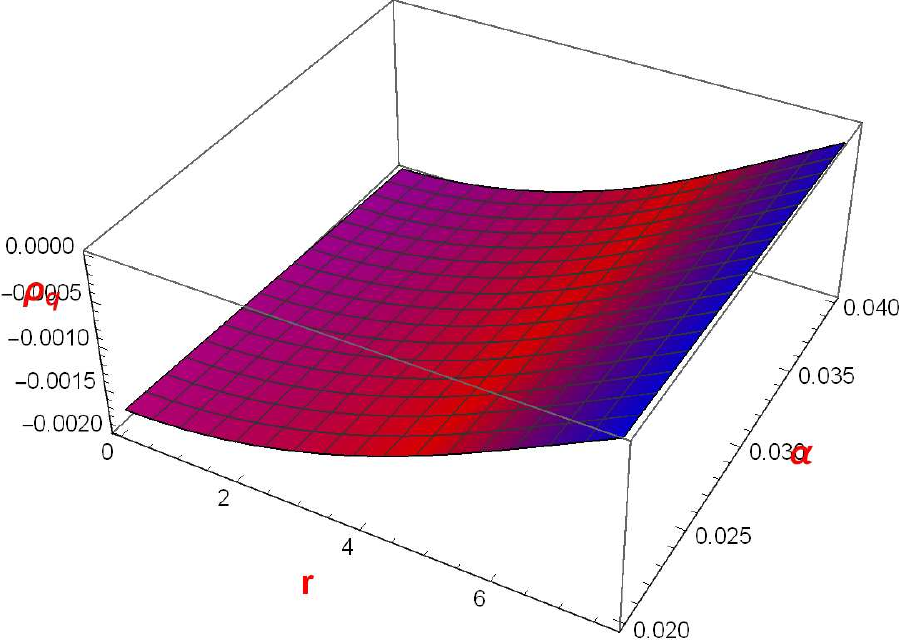, width=.32\linewidth,
height=2.5in}\epsfig{file=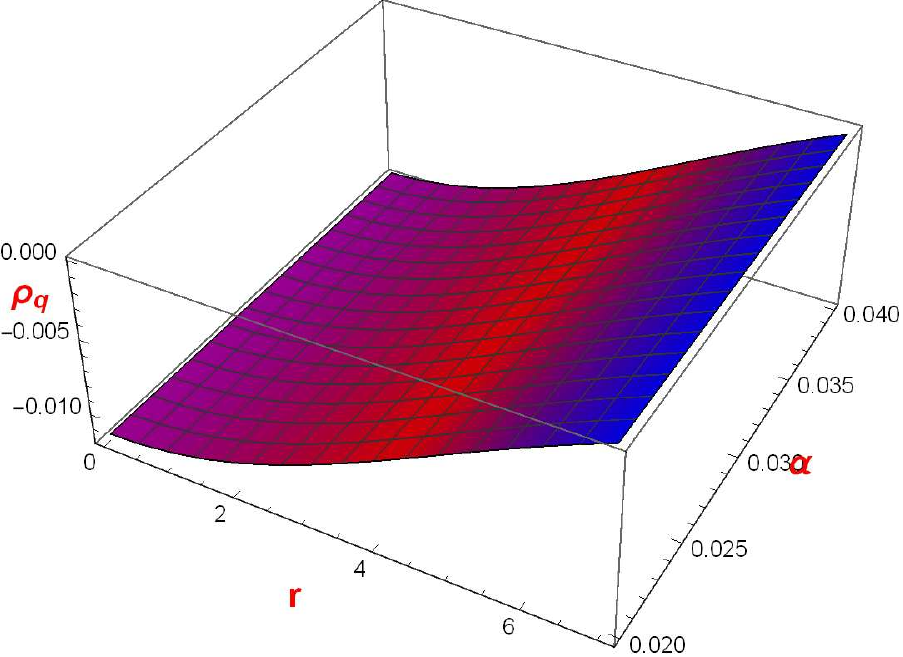, width=.32\linewidth,
height=2.5in}\epsfig{file=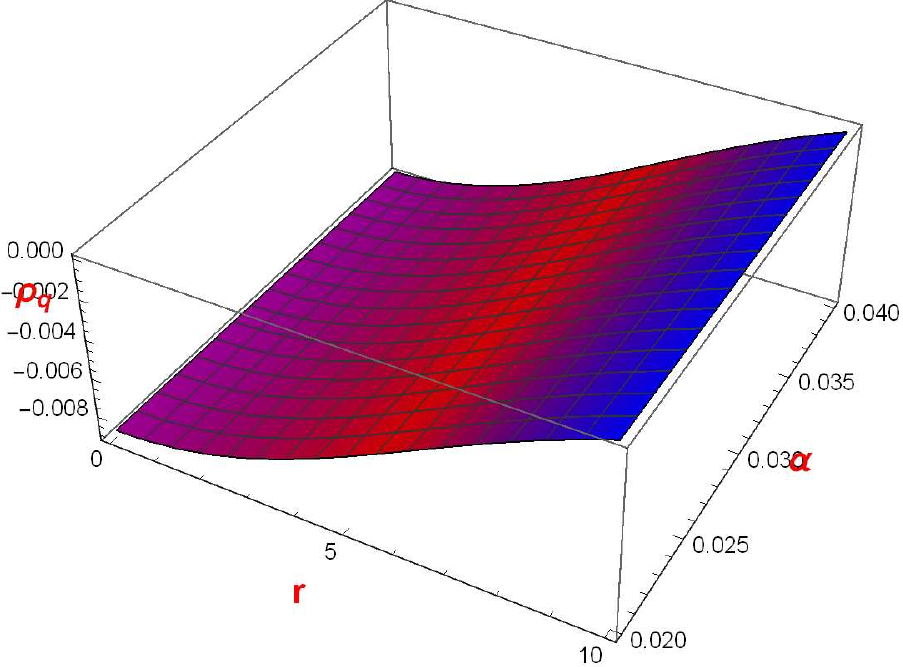, width=.32\linewidth,
height=2.5in}\caption{\label{Fig.4} shows the behavior of quintessence energy density.}
\end{figure}

Here, we discuss the energy-momentum components for the compact star solutions under the quintessence field. To have a realistic interior solution of the compact star, the energy density, transverse, and radial pressure needed to be positive. Furthermore, all of the energy-momentum components should have finite values at the center of the compact star. Figures \ref{Fig.1}, \ref{Fig.2}, \ref{Fig.3} and \ref{Fig.4} show the portraits of the energy density,  radial pressure, transverse pressure, and quintessence energy density, respectively, for HerX-1, SAXJ1808.4-3658 and 4U1820-30 stars. From those Figures, it is observed that all of the energy-momentum components gradually decrease from the center to the surface of the compact star. But, the quintessence energy density converges to zero towards the surface. Furthermore, all of the above quantities have finite values at the center of the stars (i.e., at $r=0$).

\subsection{Equation of state parameters (EoS)}\label{sec4b}
The EoS for a natural compact star was derived by Das et al. \cite{Das/2019} is a linear function of radial component $r$. However, in our work, the EoS parameters are nonlinear. The EoS for radial and transverse pressure derived in the following:
\begin{align}
\omega_r=\frac{p_r}{\rho},\,\,\ \omega_t=\frac{p_t}{\rho}.
\end{align}

\begin{figure}[H]
\centering \epsfig{file=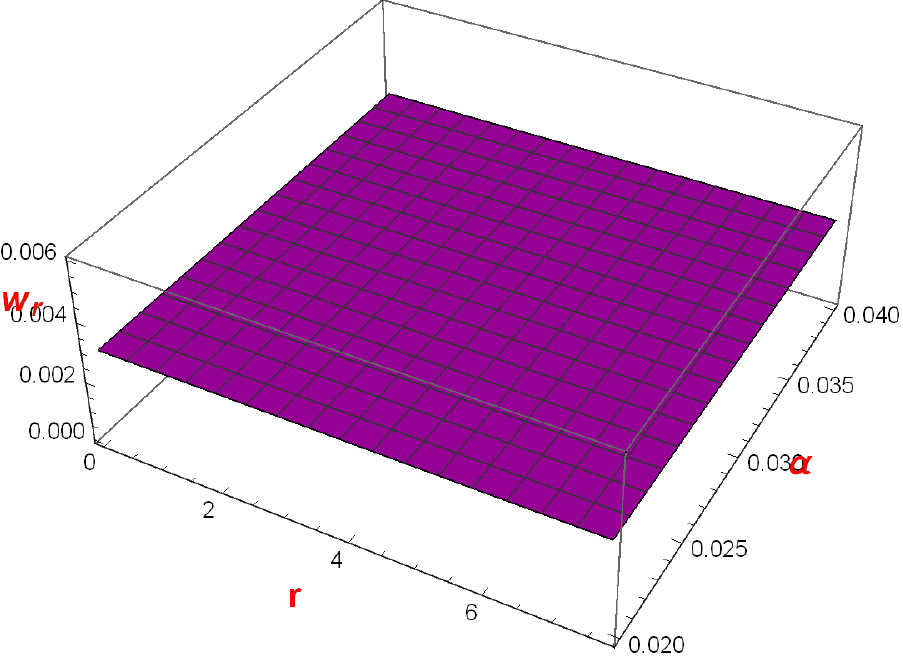, width=.32\linewidth,
height=2.5in}\epsfig{file=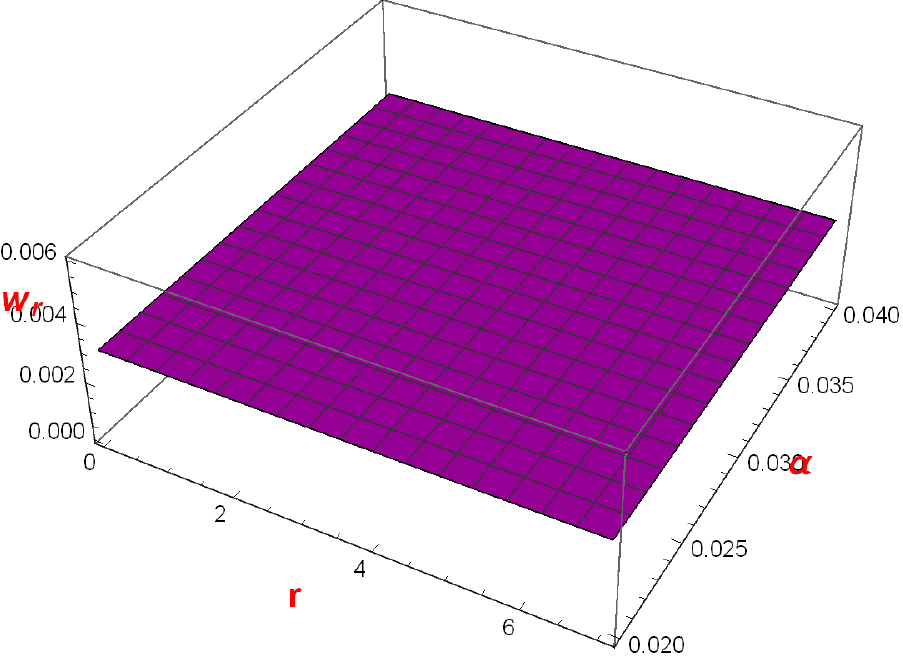, width=.32\linewidth,
height=2.5in}\epsfig{file=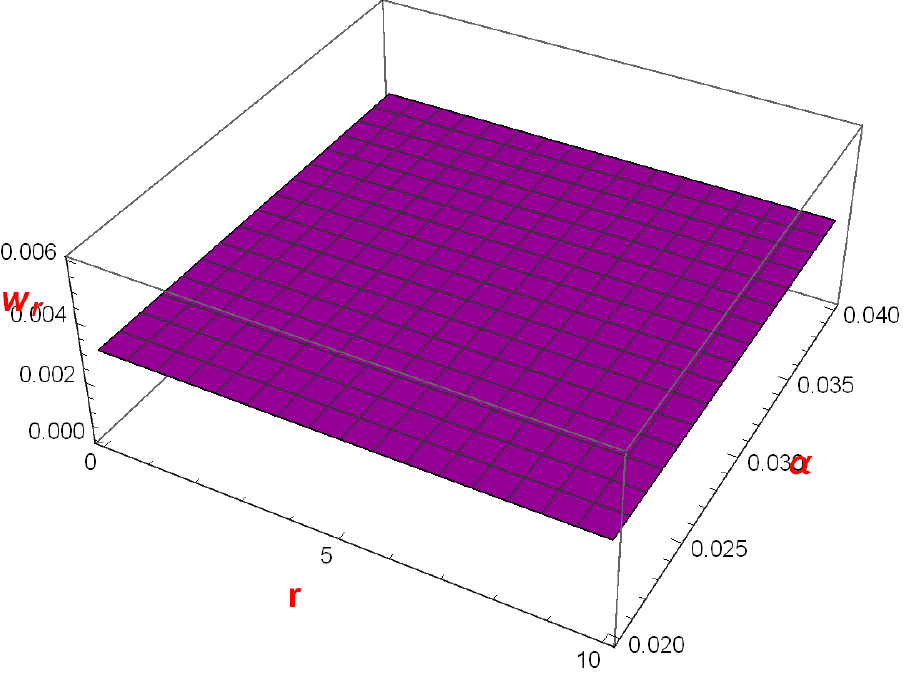, width=.32\linewidth,
height=2.5in}\caption{\label{Fig.5} shows the behavior of $w_{r}$.}
\end{figure}

\begin{figure}[H]
\centering \epsfig{file=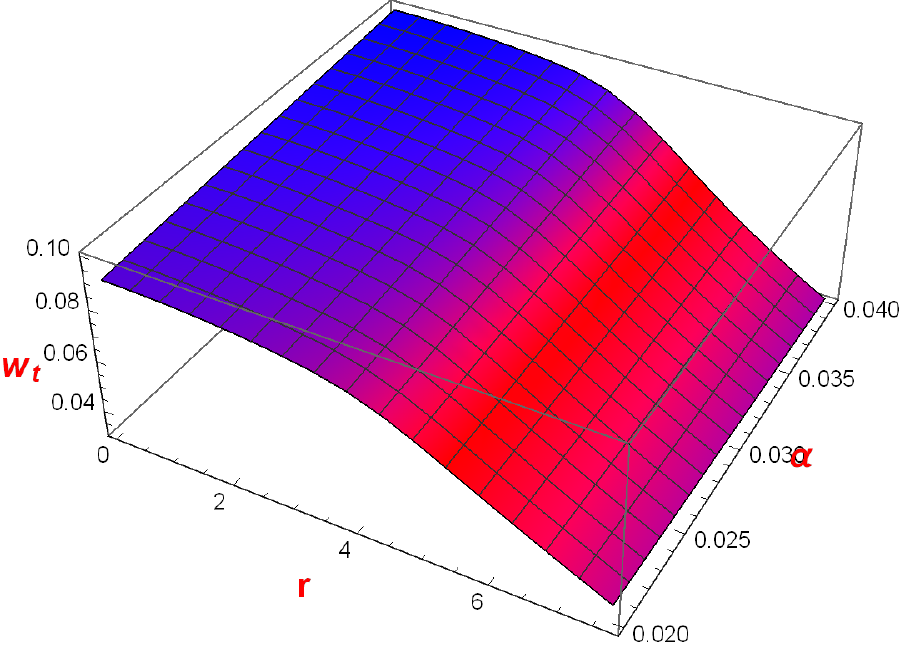, width=.32\linewidth,
height=2.5in}\epsfig{file=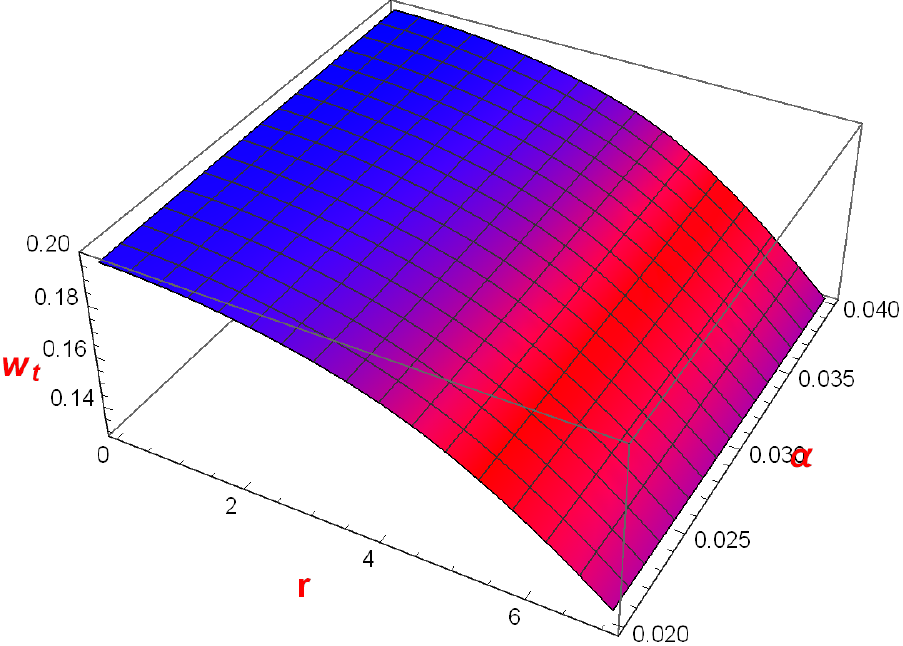, width=.32\linewidth,
height=2.5in}\epsfig{file=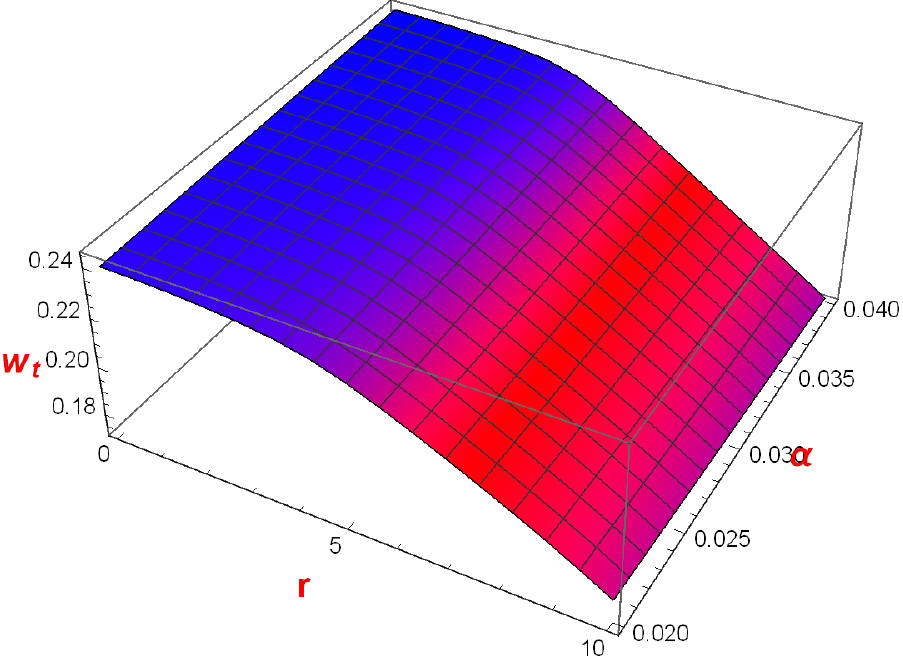, width=.32\linewidth,
height=2.5in}\caption{\label{Fig.6} shows the behavior of $w_{t}$.}
\end{figure}

Figures \ref{Fig.5} and \ref{Fig.6} depict the profiles of the EoS for the radial and transverse pressure, respectively. It is seen that both of the important ratios lie between 0 and 1 for three compact star models. Thus, all discussed solutions are physically acceptable in the framework of $f(Q)$ gravity.

\subsection{Gradients}\label{sec4c}
Now we examine the gradients of the energy density, radial, and tangential pressure for our compact star models. To construct a real compact star model, gradients have to satisfy the following conditions \cite{Chanda/2019}
\begin{align}
\frac{d\rho}{dr}<0,\,\,\ \frac{dp_r}{dr}<0,\,\,\ \frac{dp_t}{dr}<0.
\end{align}

\begin{figure}[H]
\centering \epsfig{file=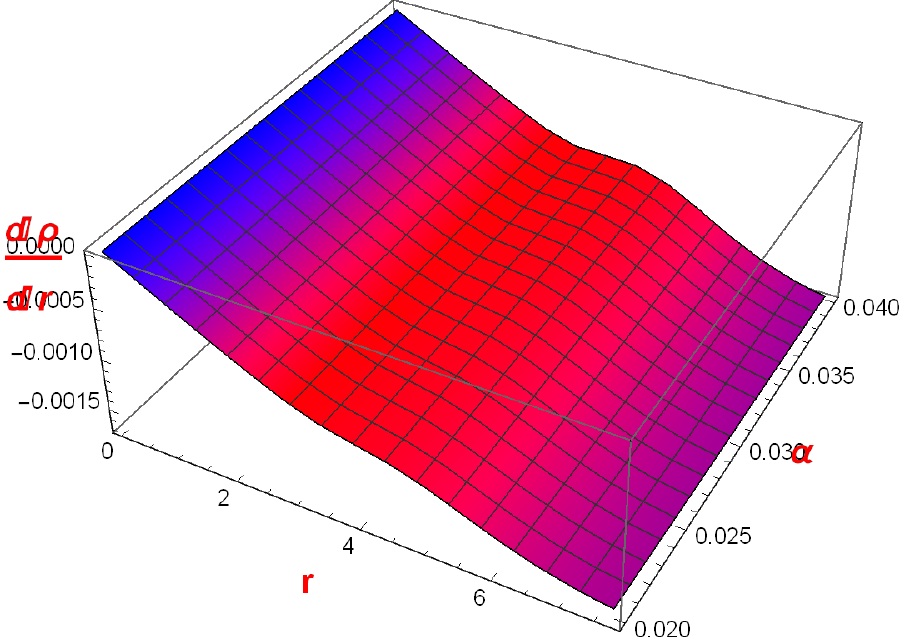, width=.32\linewidth,
height=2.5in}\epsfig{file=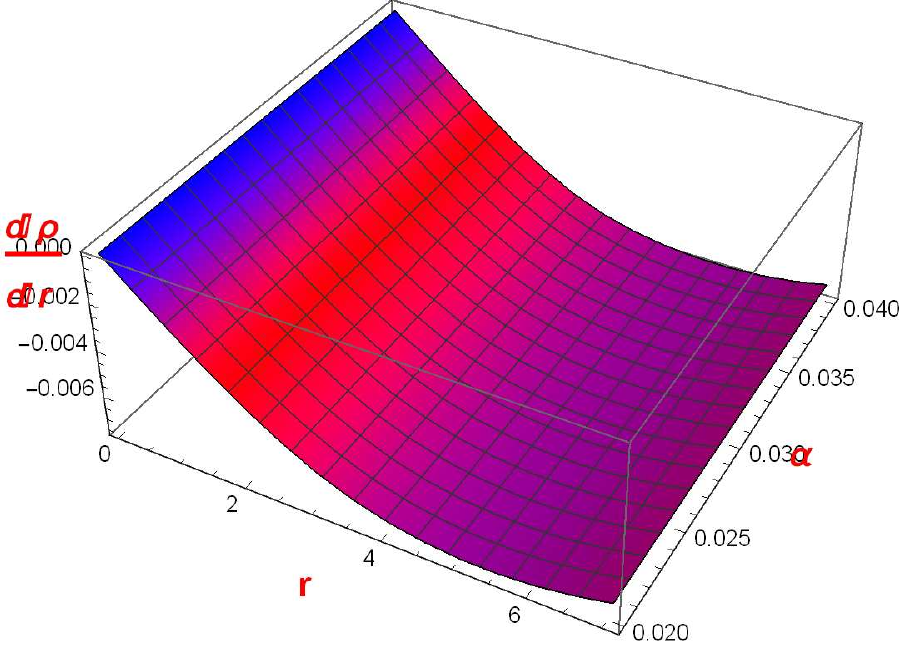, width=.32\linewidth,
height=2.5in}\epsfig{file=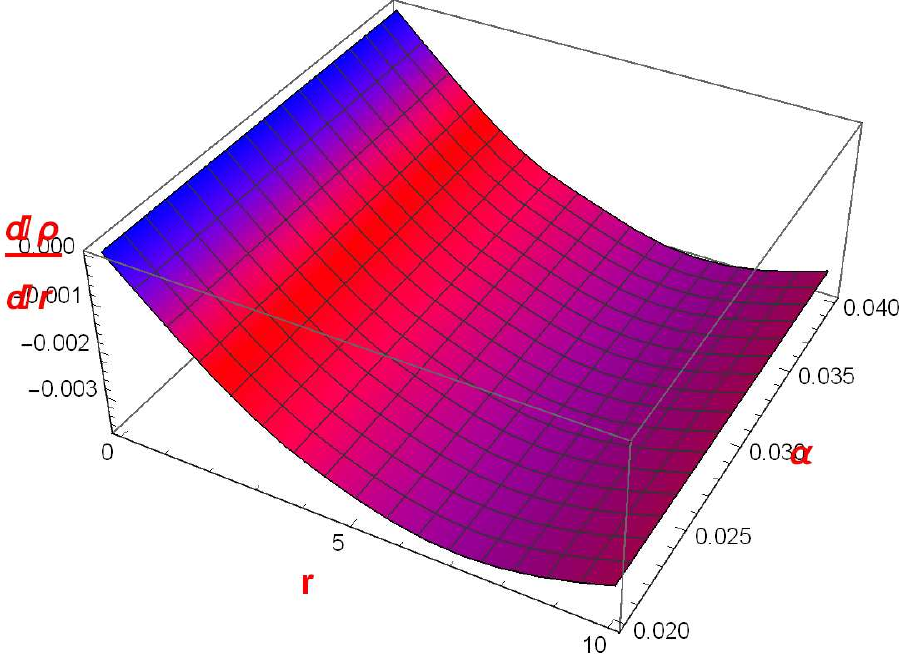, width=.32\linewidth,
height=2.5in}\caption{\label{Fig.7} shows the behavior of $\frac{d\rho}{dr}$.}
\end{figure}

\begin{figure}[H]
\centering \epsfig{file=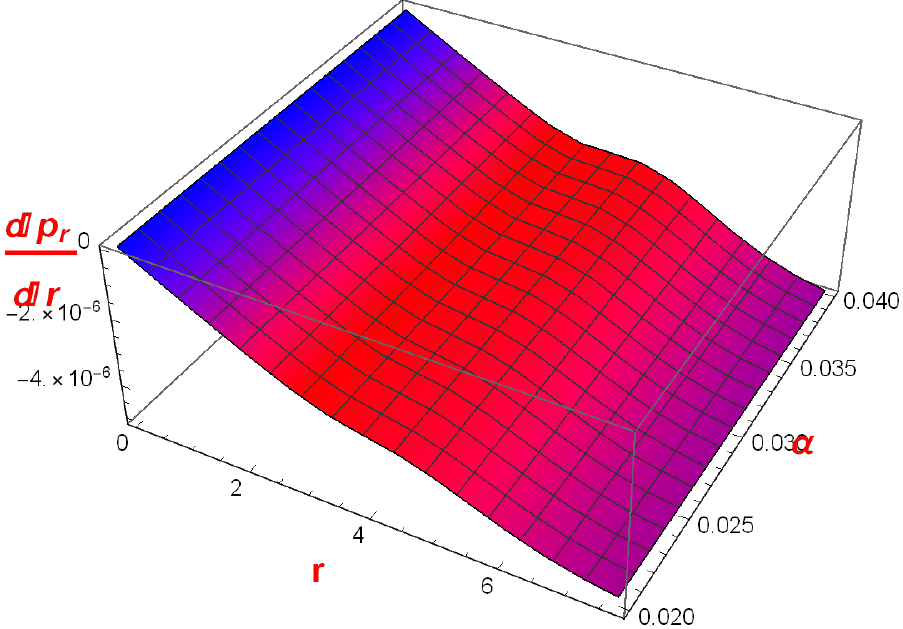, width=.32\linewidth,
height=2.5in}\epsfig{file=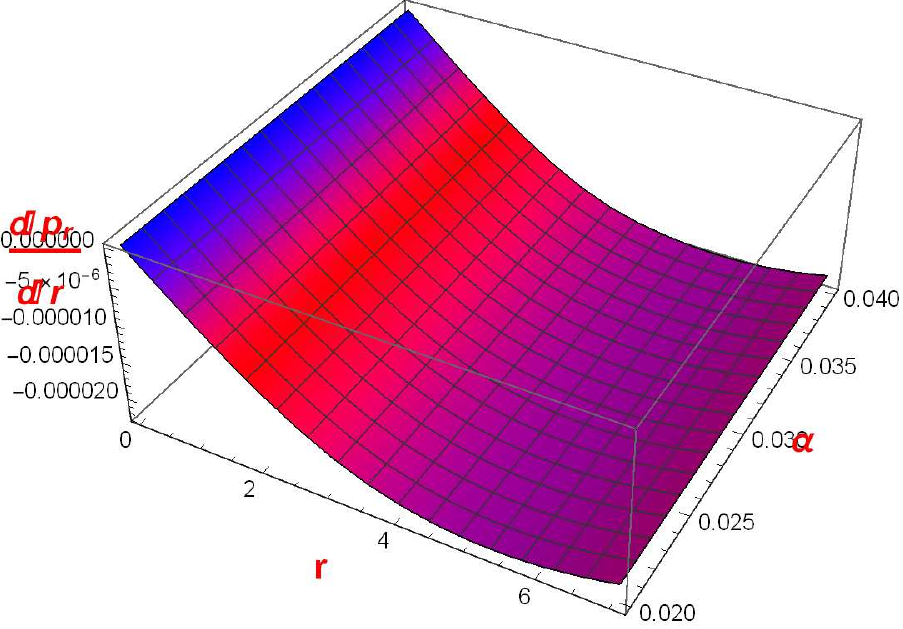, width=.32\linewidth,
height=2.5in}\epsfig{file=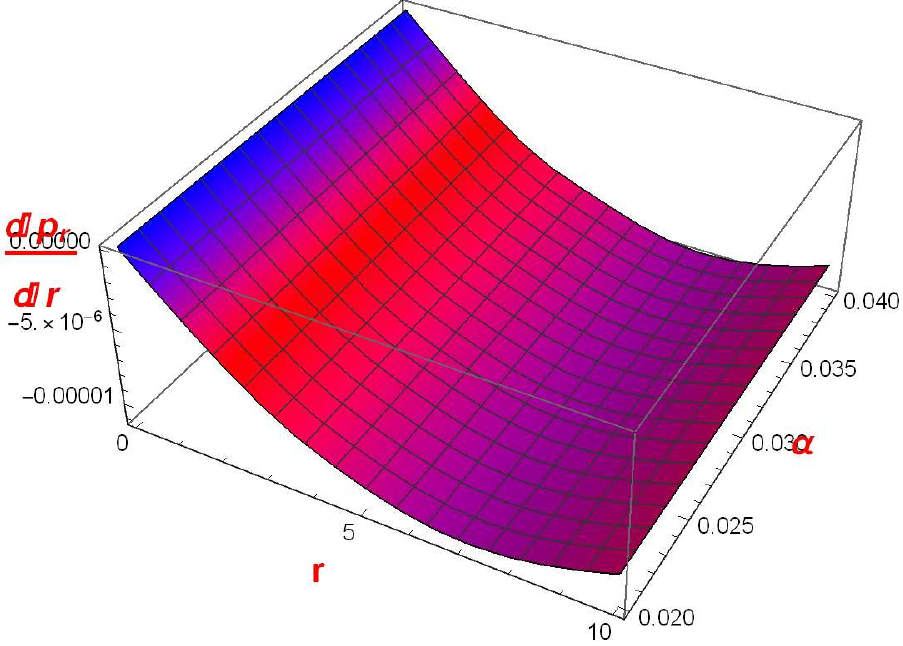, width=.32\linewidth,
height=2.5in}\caption{\label{Fig.8} shows the behavior of $\frac{dp_{r}}{dr}$.}
\end{figure}

\begin{figure}[H]
\centering \epsfig{file=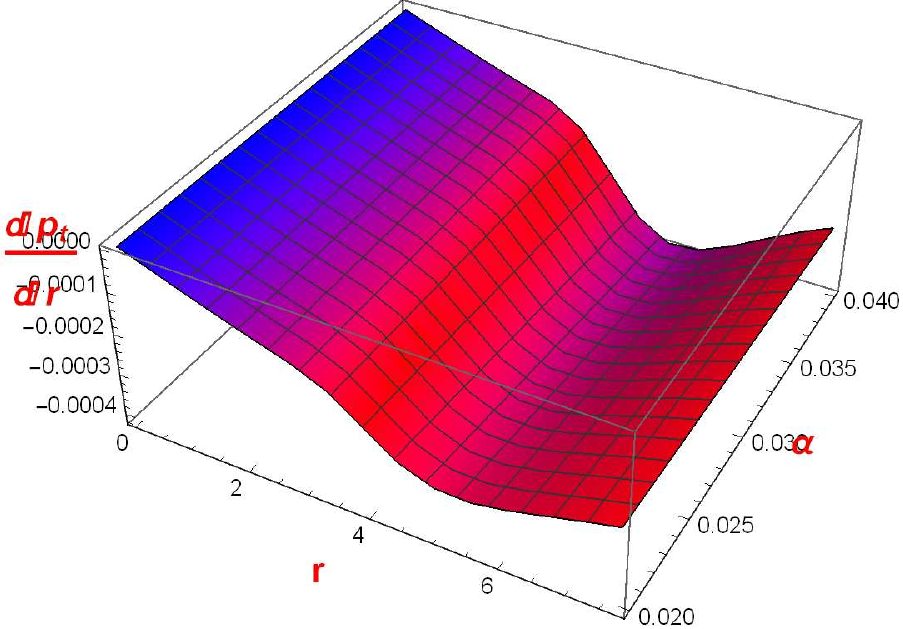, width=.32\linewidth,
height=2.5in}\epsfig{file=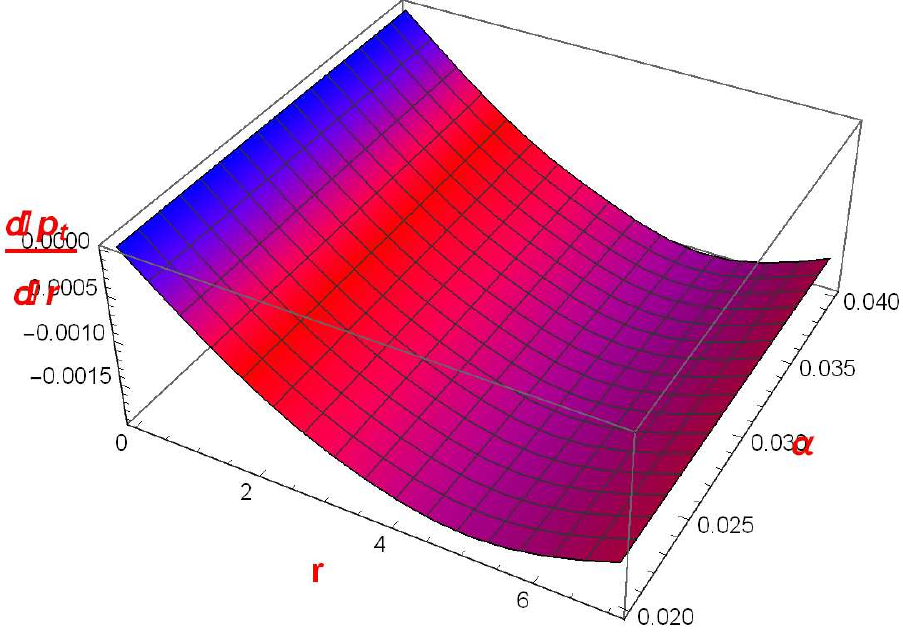, width=.32\linewidth,
height=2.5in}\epsfig{file=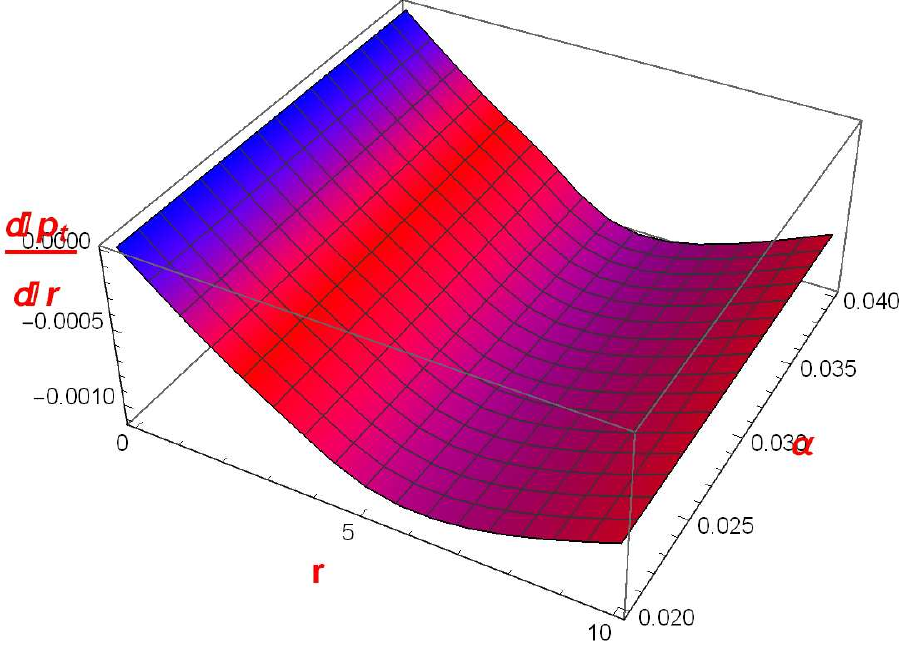, width=.32\linewidth,
height=2.5in}\caption{\label{Fig.9} shows the behavior of $\frac{dp_{t}}{dr}$.}
\end{figure}

The above conditions on gradients can be verified from Figures \ref{Fig.7}, \ref{Fig.8}, and \ref{Fig.9}. The negative behavior of gradients is the necessary condition for a realistic compact star model. In addition, the gradients are vanished at $r=0$ i.e.,

\begin{align}
\frac{d\rho}{dr}\bigg|_{r=0}=0,\,\,\ \frac{dp_r}{dr}\bigg|_{r=0}=0,\,\,\ \frac{dp_t}{dr}\bigg|_{r=0}=0.
\end{align}

\subsection{Anisotropy}\label{sec4d}

\begin{figure}[H]
\centering \epsfig{file=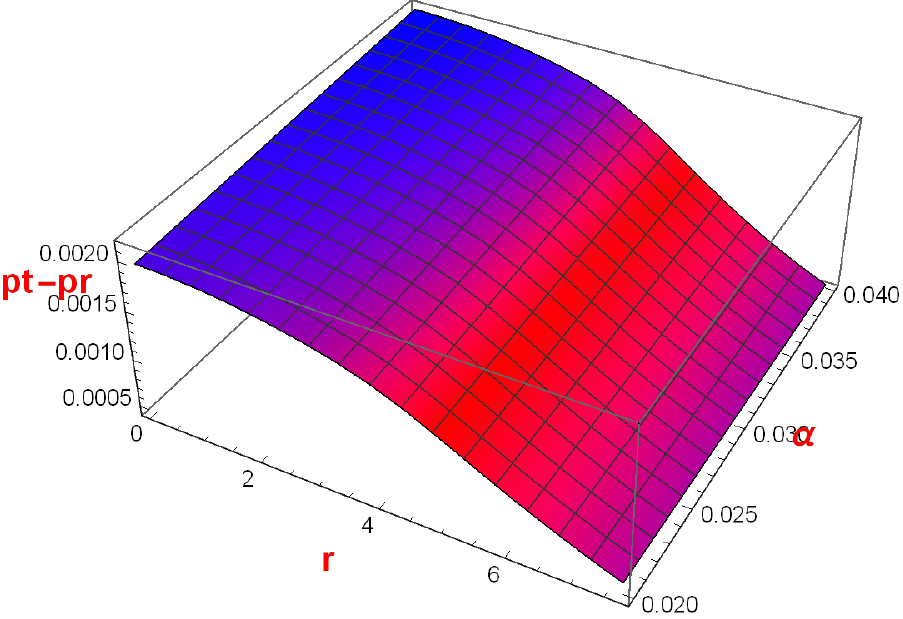, width=.32\linewidth,
height=2.5in}\epsfig{file=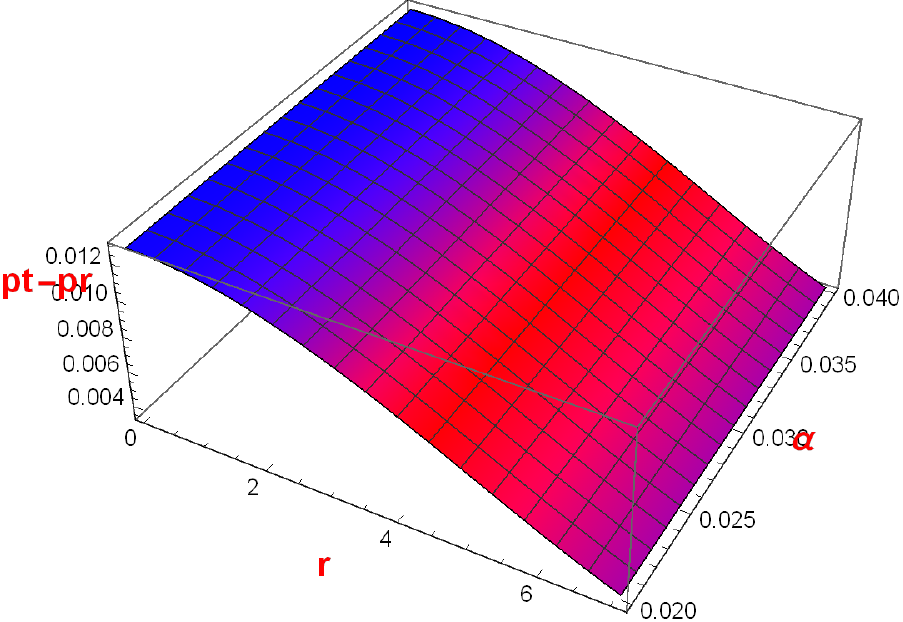, width=.32\linewidth,
height=2.5in}\epsfig{file=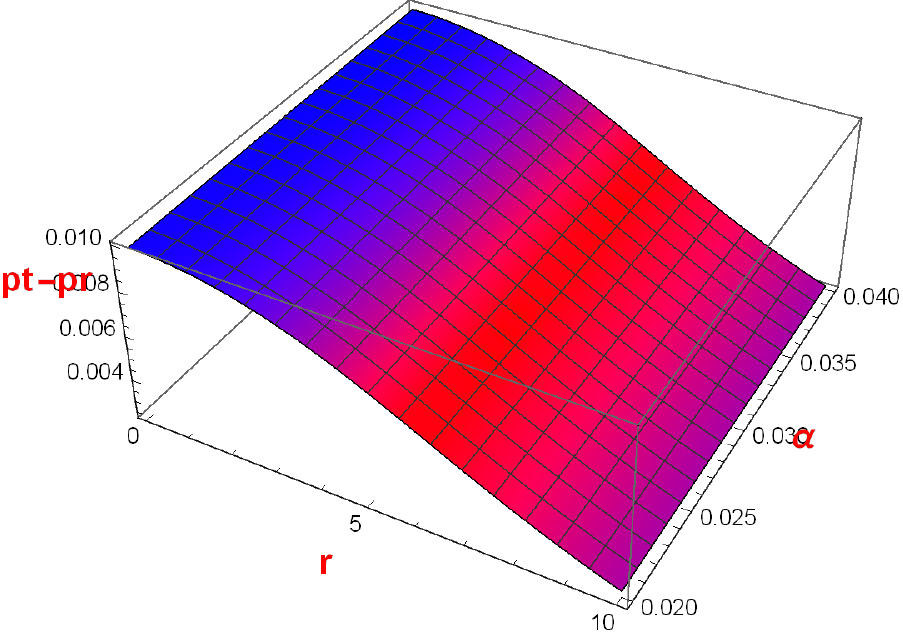, width=.32\linewidth,
height=2.5in}\caption{\label{Fig.10} shows the behavior of Anisotropy function.}
\end{figure}

For massive stellar objects, the radial and traverse pressures may not be equal. Nevertheless, several proposals have been given for the existence of stellar anisotropy structures, such as different kinds of phase transitions \cite{Sokoloz/1980} and the presence of type 3A superfluid \cite{Kippenhahm/1990}. Furthermore, the anisotropy helps to understand the strange properties of matter at the core of the compact star. The anisotropy function can be defined as the difference between the pressure components, i.e., $\Delta=p_t-p_r$. Figure  \ref{Fig.10} displays the profile of the anisotropy function and reveals that the anisotropy force is positive. This result suggests that $p_t>p_r$ and anisotropic force is repulsive.

\subsection{Causality and Abreu stability analysis}\label{sec4e}
In this subsection, we check the stability of our models using the well-known causality condition. We test two types of speeds of sound, i.e., radial and tangential speeds of sounds, which can be defined by $v_r^2$ and $v_t^2$, calculated as follows:

\begin{align}
v_r^2=\frac{dp_r}{d\rho},\,\,\ v_t^2=\frac{dp_t}{d\rho}.
\end{align}

\begin{figure}[H]
\centering \epsfig{file=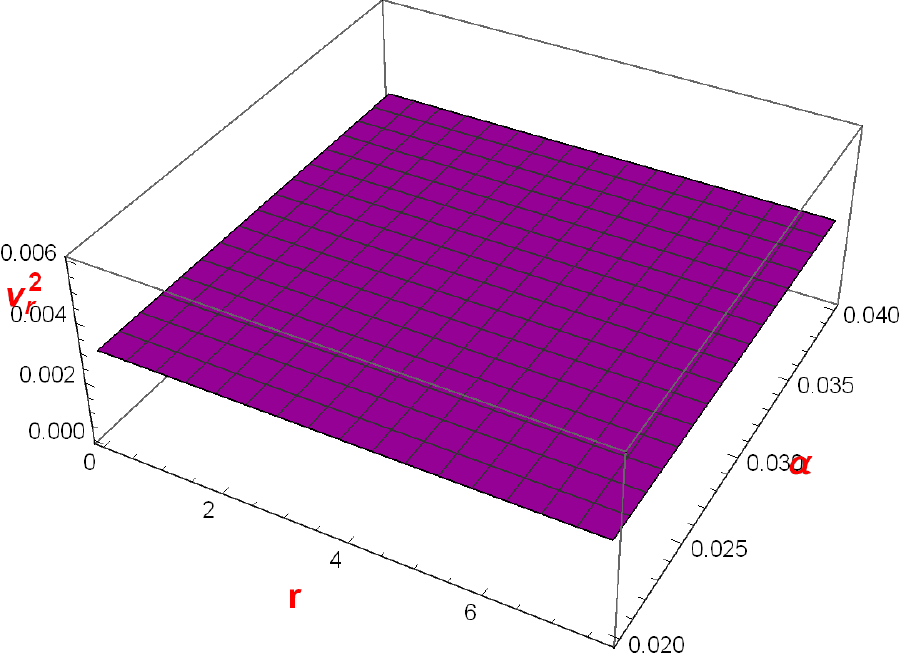, width=.32\linewidth,
height=2.5in}\epsfig{file=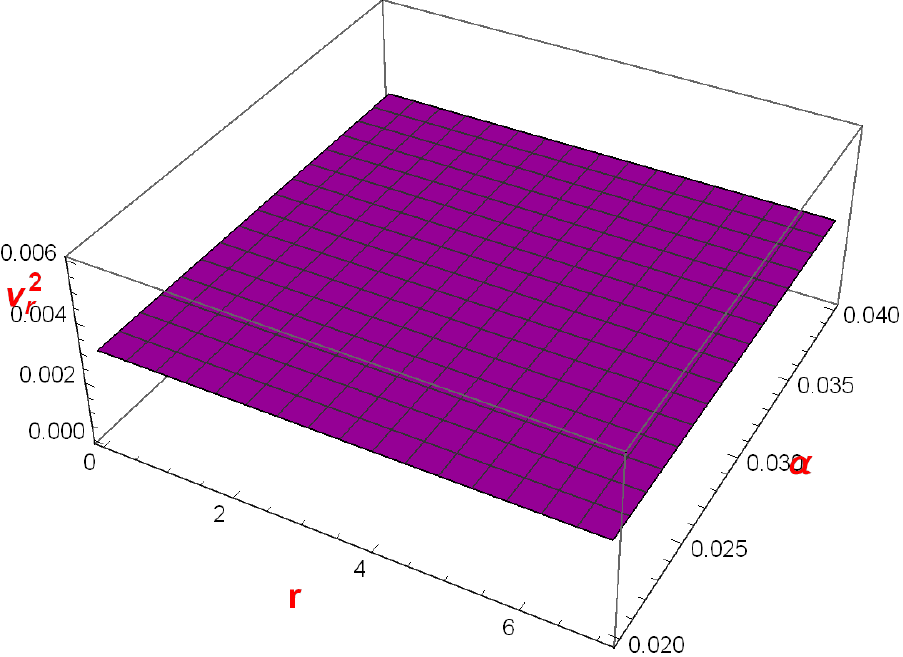, width=.32\linewidth,
height=2.5in}\epsfig{file=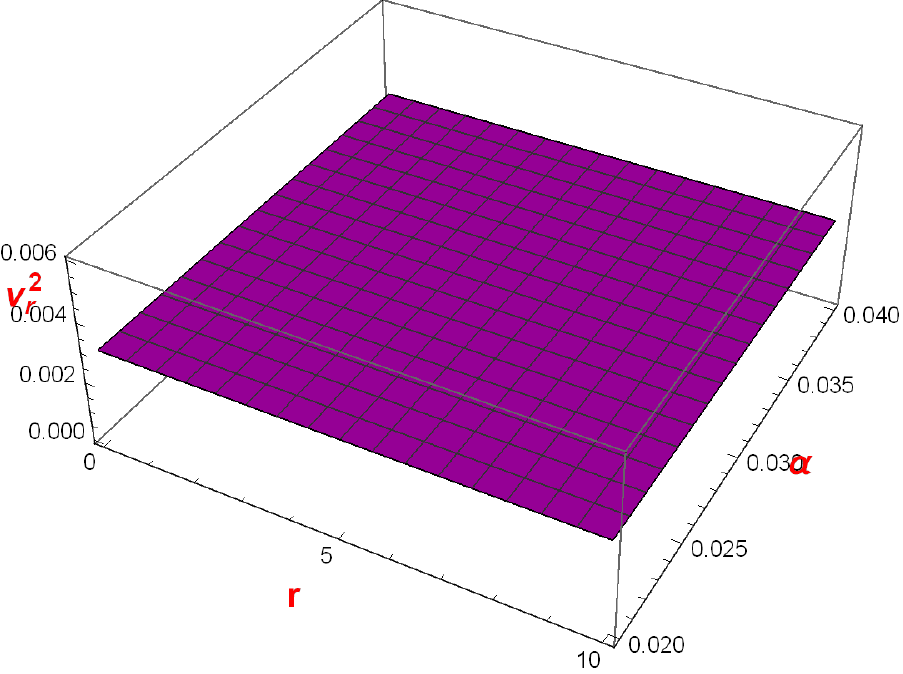, width=.32\linewidth,
height=2.5in}\caption{\label{Fig.11} shows the behavior of $v^{2}_{r}$.}
\end{figure}

\begin{figure}[H]
\centering \epsfig{file=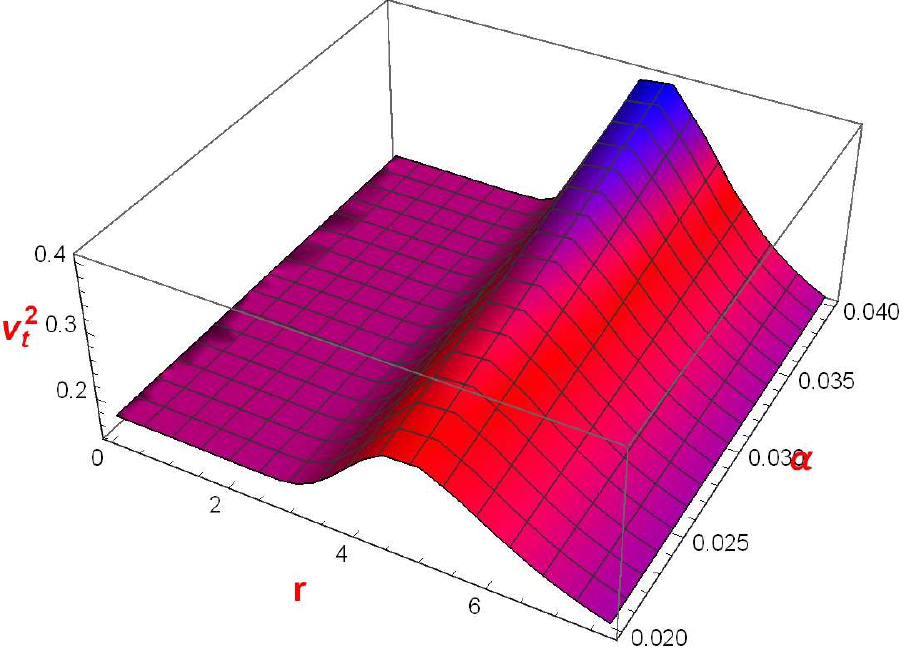, width=.32\linewidth,
height=2.5in}\epsfig{file=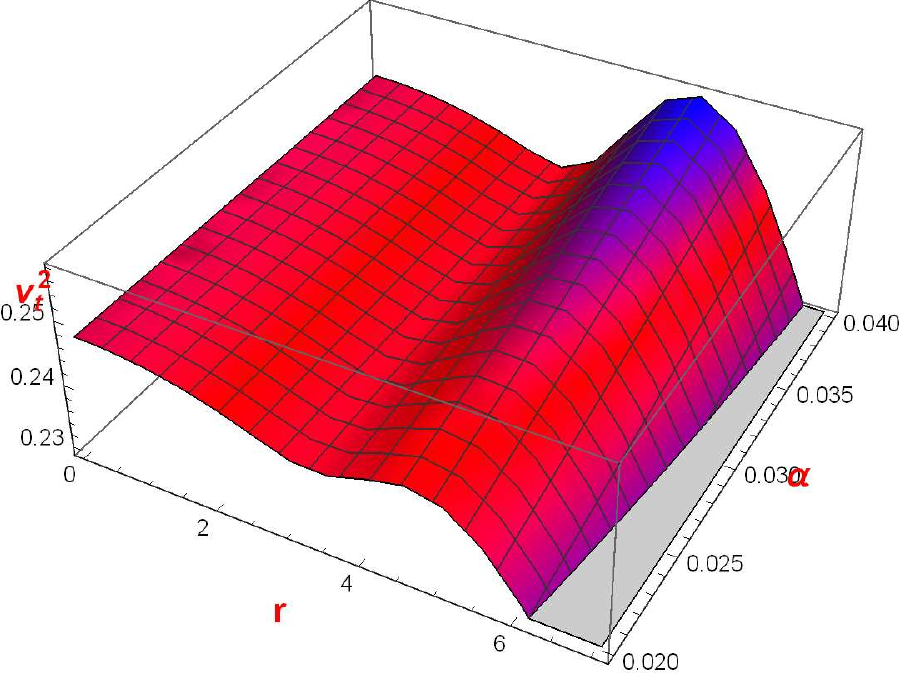, width=.32\linewidth,
height=2.5in}\epsfig{file=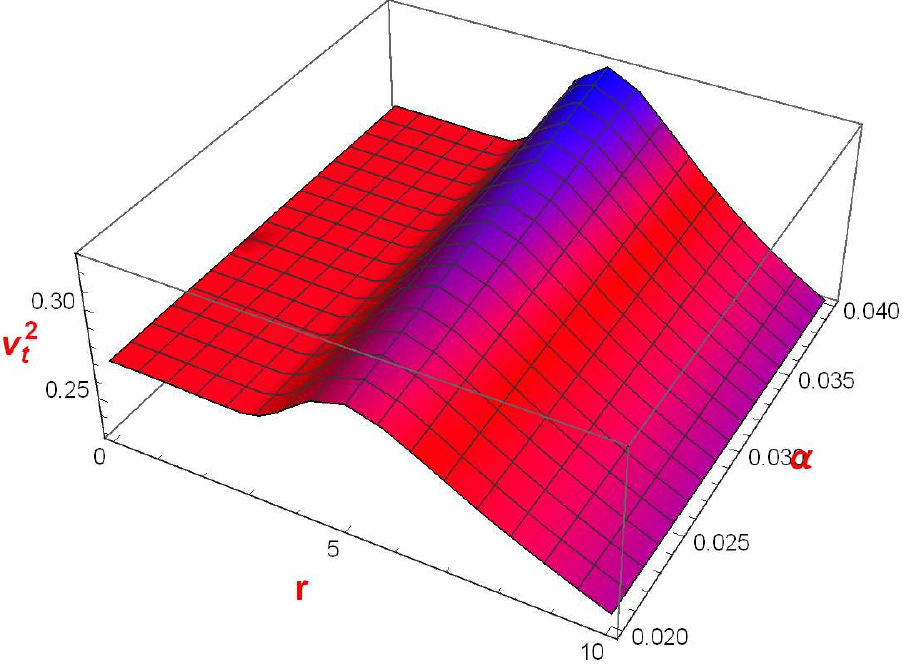, width=.32\linewidth,
height=2.5in}\caption{\label{Fig.12} shows the behavior of $v^{2}_{t}$.}
\end{figure}

\begin{figure}[H]
\centering \epsfig{file=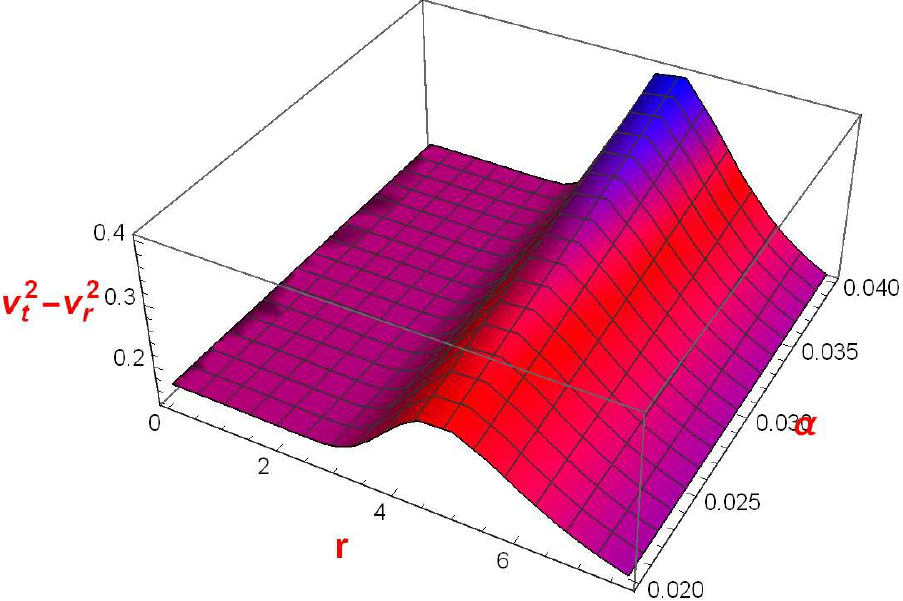, width=.32\linewidth,
height=2.5in}\epsfig{file=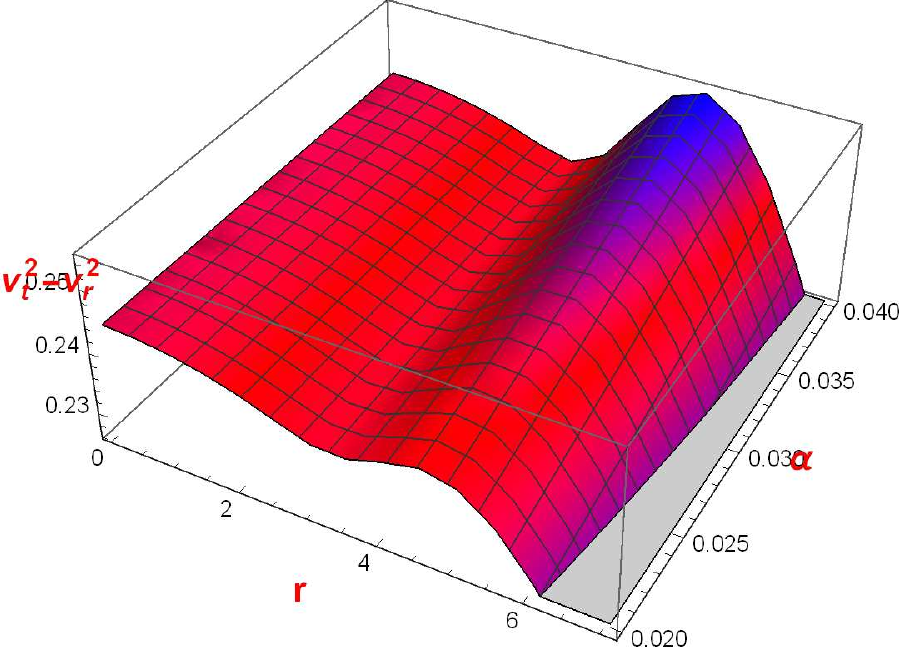, width=.32\linewidth,
height=2.5in}\epsfig{file=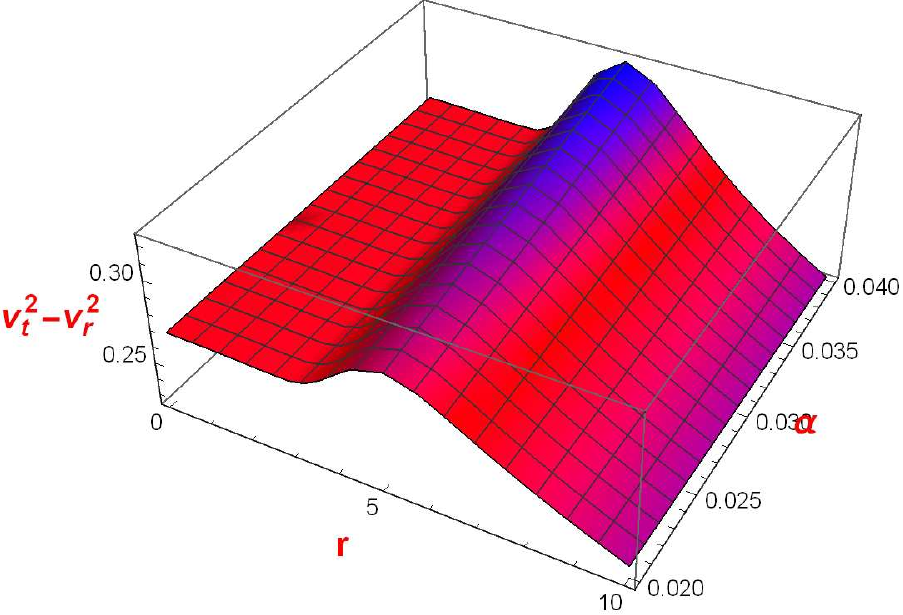, width=.32\linewidth,
height=2.5in}\caption{\label{Fig.13} shows the behavior of $v^{2}_{t}-v^{2}_{r}$.}
\end{figure}

\begin{figure}[H]
\centering \epsfig{file=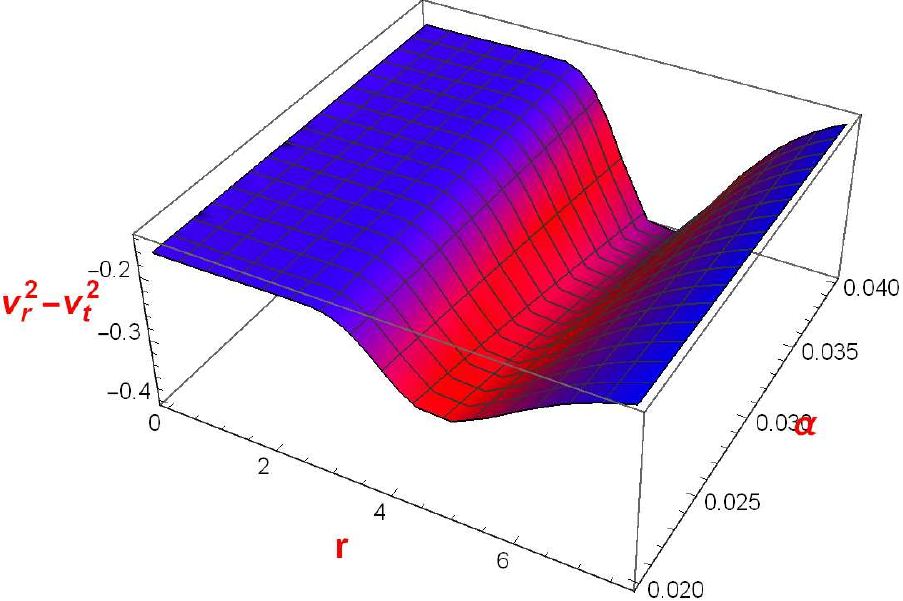, width=.32\linewidth,
height=2.5in}\epsfig{file=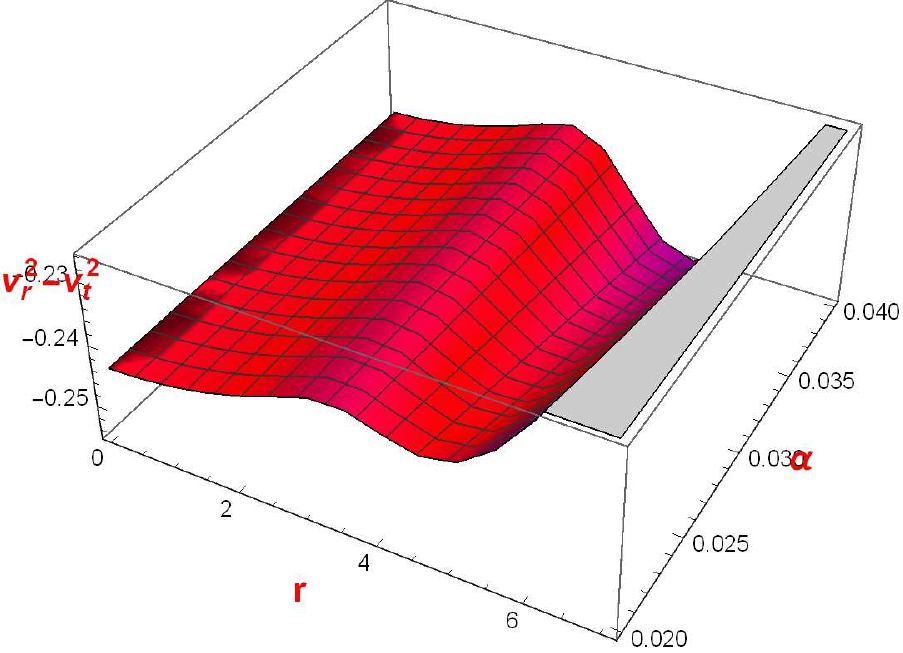, width=.32\linewidth,
height=2.5in}\epsfig{file=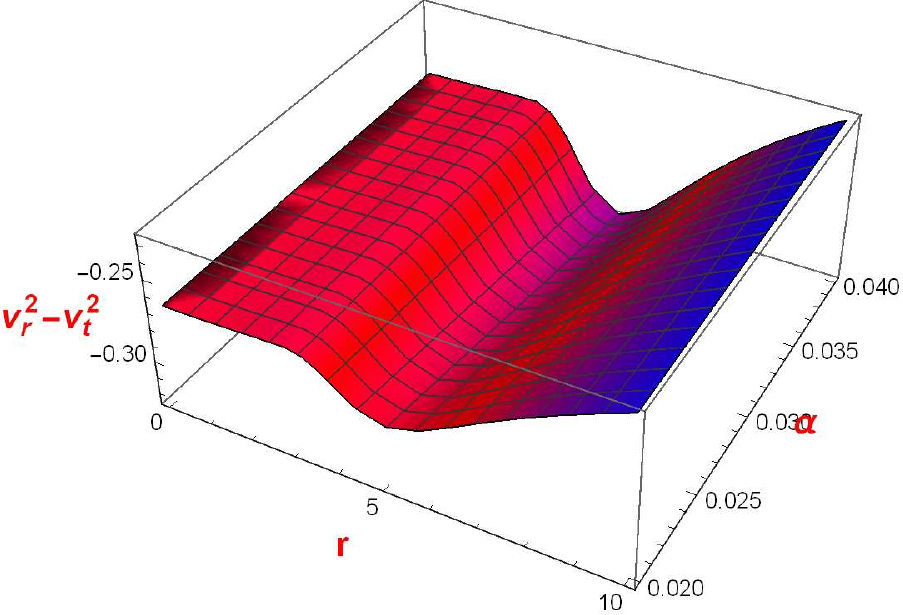, width=.32\linewidth,
height=2.5in}\caption{\label{Fig.14} shows the behavior of $v^{2}_{r}-v^{2}_{t}$.}
\end{figure}

The radial sound speed $v_r^2$ and tangential sound speed $v_t^2$ should be less than the speed of light to confirm that the causality condition is satisfied. To ensure that, we present the sound speeds in Figures \ref{Fig.11} and \ref{Fig.12}. It is seen that the sound speeds are less than the speed of light (considering the speed of light is unity in relativistic units). Furthermore, we have tested the cracking condition to avoid gravitational cracking for a stable anisotropic compact star. The cracking condition or Abreu condition i.e., $-1\leq v_t^2-v_r^2\leq 0$ illustrates in \ref{Fig.14}. It can also be observed from the
in Figure \ref{Fig.13}, that the expression  $0\leq v_t^2-v_r^2\leq 1$, satisfies the Abrea condition conversely.

\subsection{Energy conditions}\label{sec4f}
\begin{figure}[H]
\centering \epsfig{file=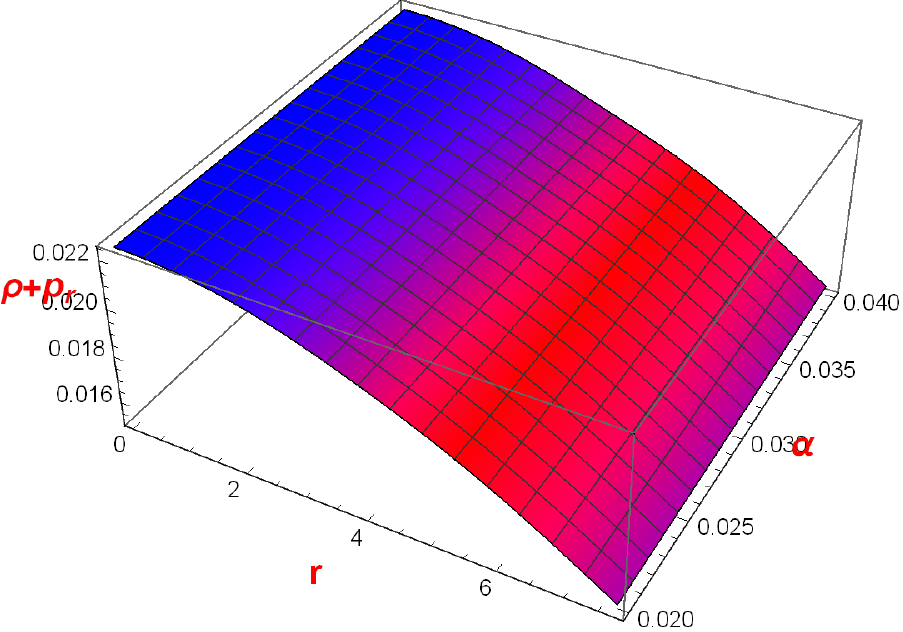, width=.32\linewidth,
height=2.5in}\epsfig{file=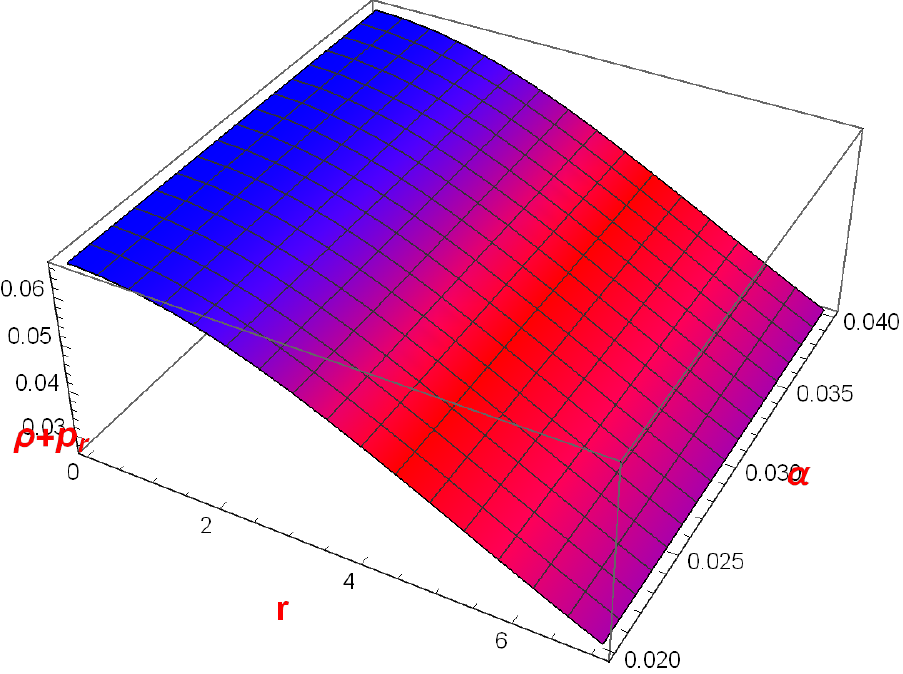, width=.32\linewidth,
height=2.5in}\epsfig{file=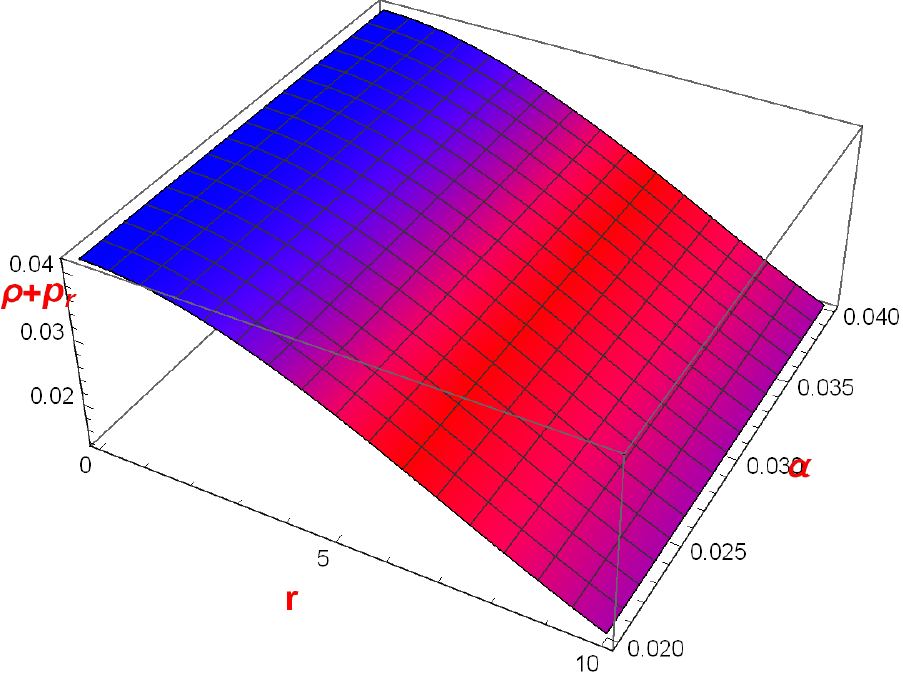, width=.32\linewidth,
height=2.5in}\caption{\label{Fig.15} shows the behavior of $\rho+p_r$.}
\end{figure}

\begin{figure}[H]
\centering \epsfig{file=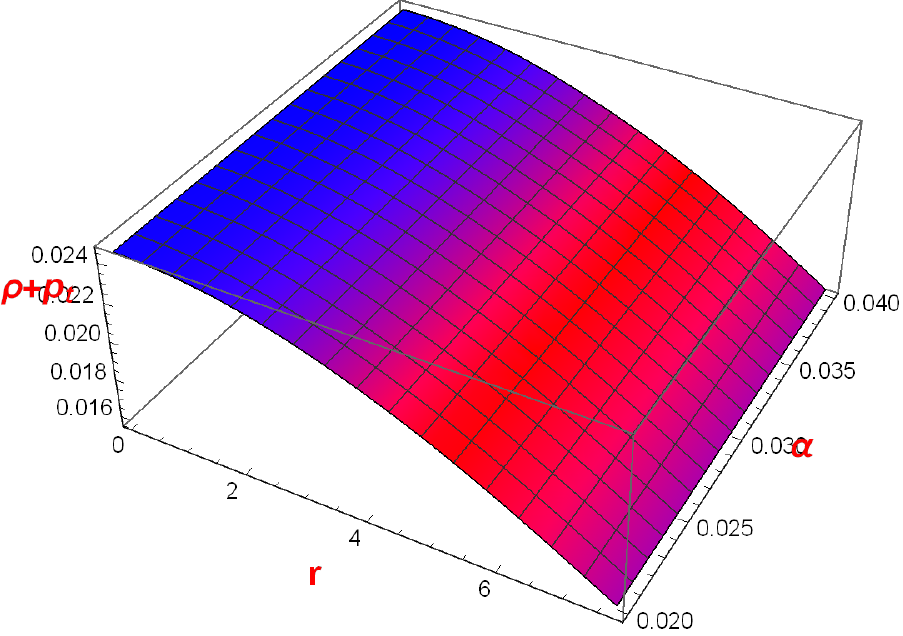, width=.32\linewidth,
height=2.5in}\epsfig{file=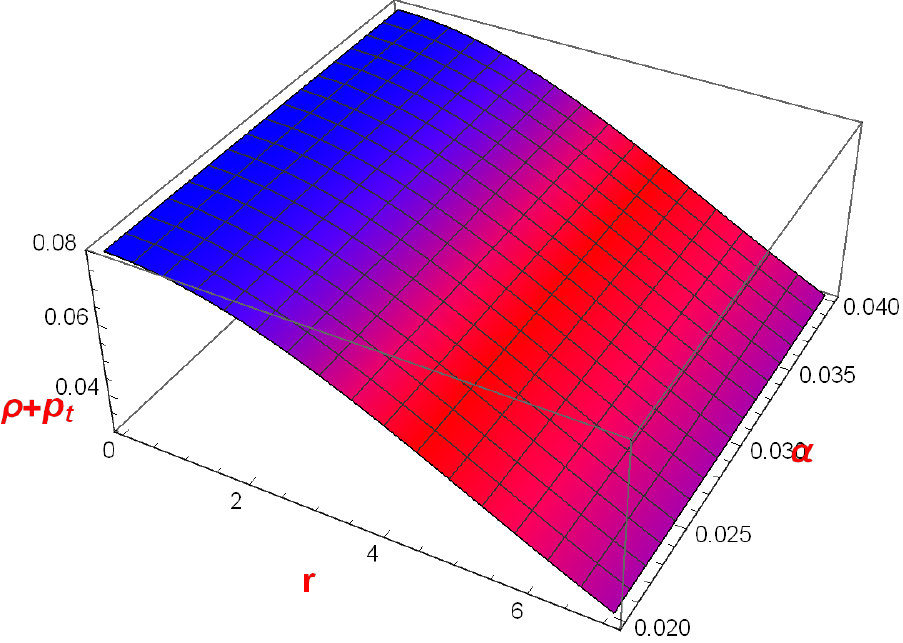, width=.32\linewidth,
height=2.5in}\epsfig{file=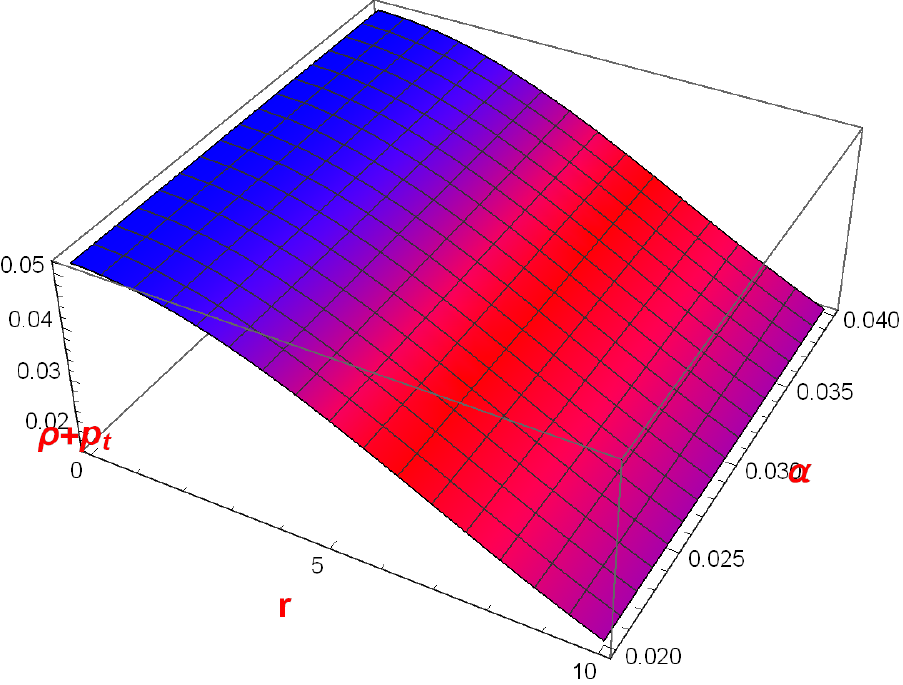, width=.32\linewidth,
height=2.5in}\caption{\label{Fig.16} shows the behavior of $\rho+p_t$.}
\end{figure}

\begin{figure}[H]
\centering \epsfig{file=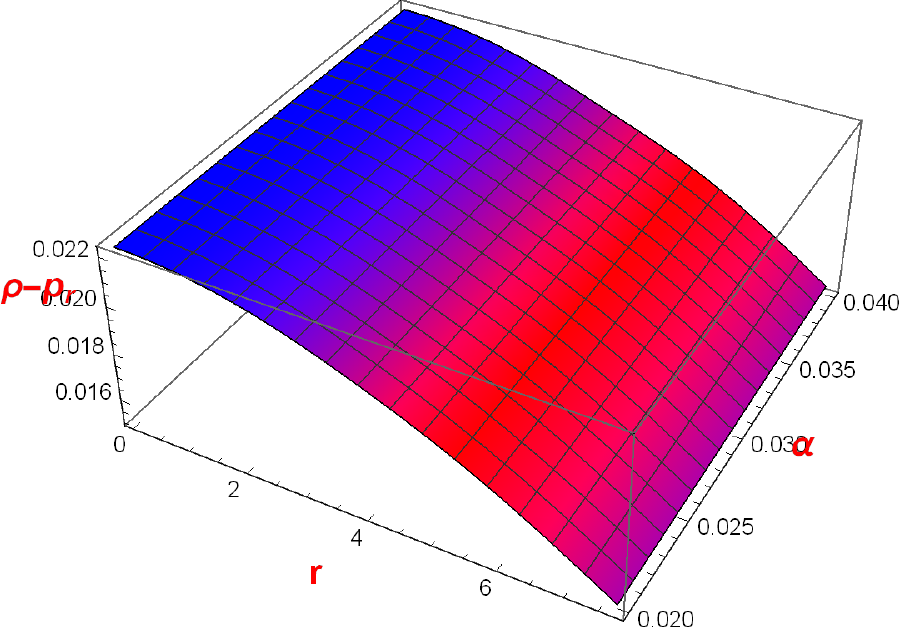, width=.32\linewidth,
height=2.5in}\epsfig{file=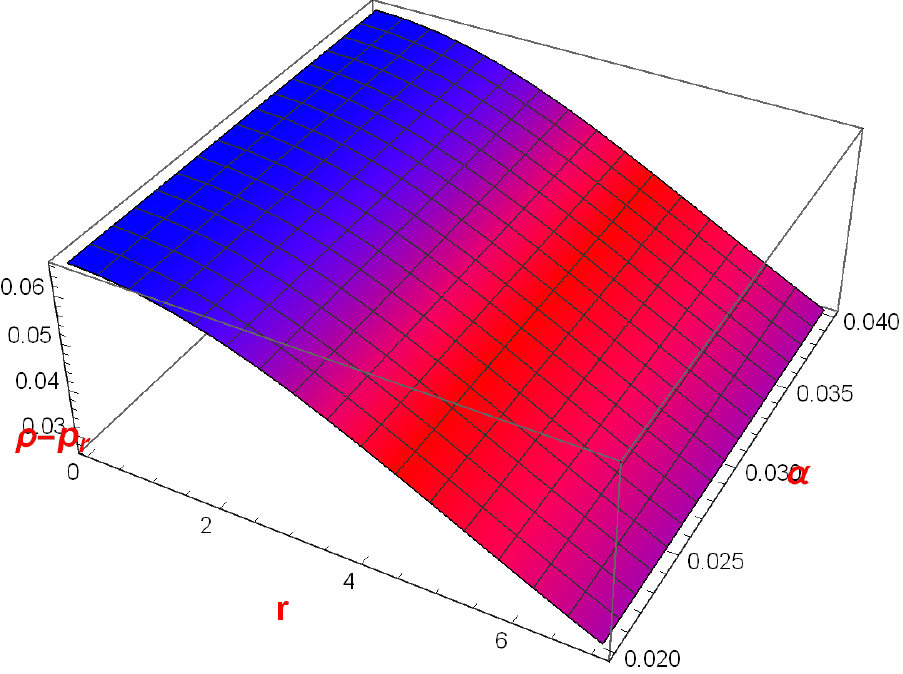, width=.32\linewidth,
height=2.5in}\epsfig{file=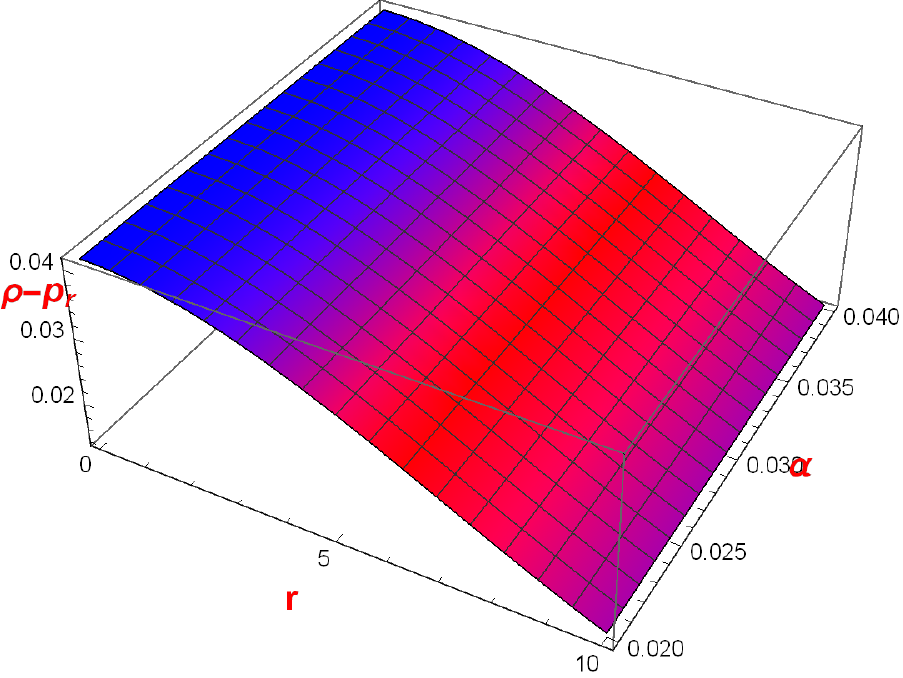, width=.32\linewidth,
height=2.5in}\caption{\label{Fig.17} shows the behavior of $\rho-p_r$.}
\end{figure}

\begin{figure}[H]
\centering \epsfig{file=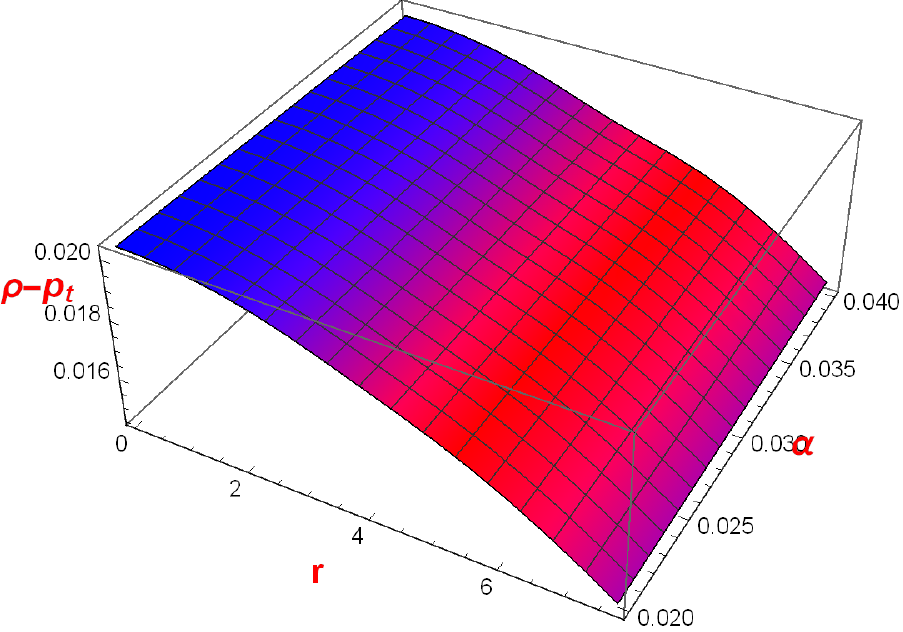, width=.32\linewidth,
height=2.5in}\epsfig{file=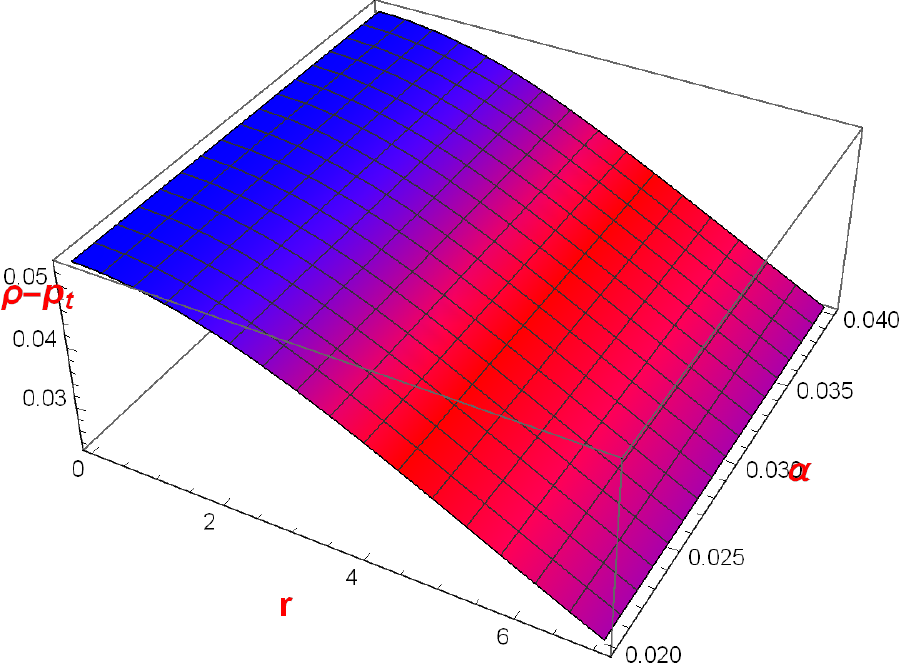, width=.32\linewidth,
height=2.5in}\epsfig{file=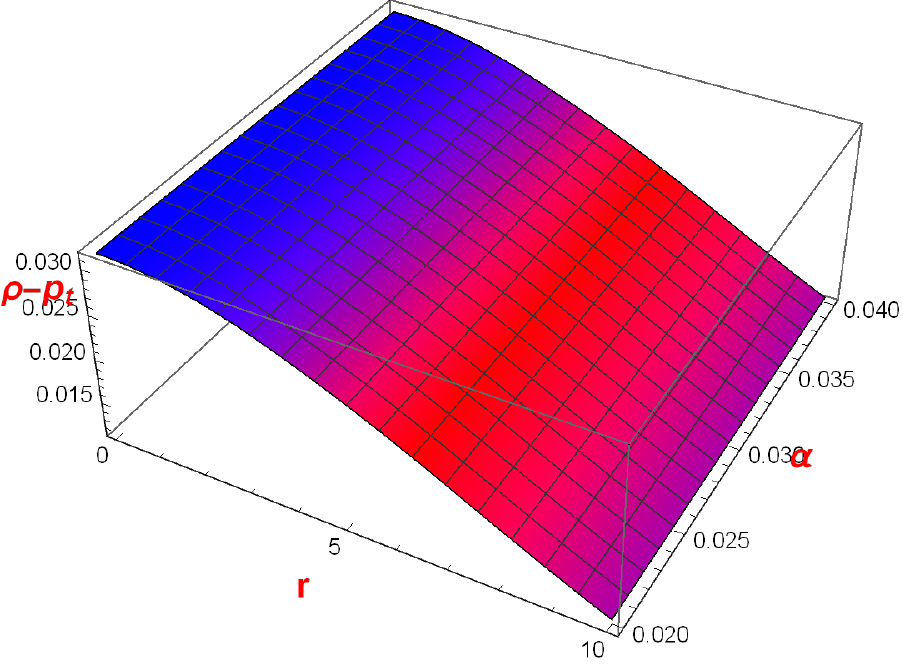, width=.32\linewidth,
height=2.5in}\caption{\label{Fig.18} shows the behavior of $\rho-p_t$.}
\end{figure}

\begin{figure}[H]
\centering \epsfig{file=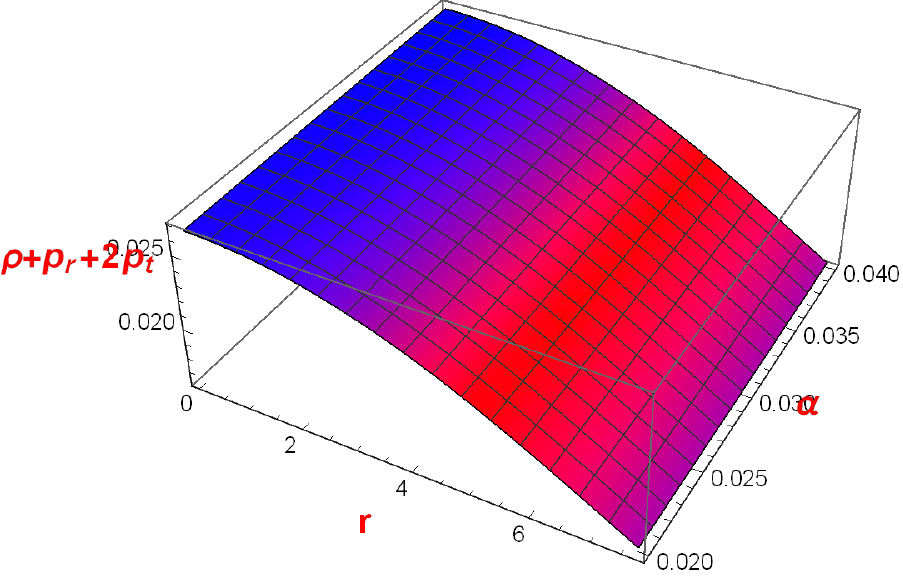, width=.32\linewidth,
height=2.5in}\epsfig{file=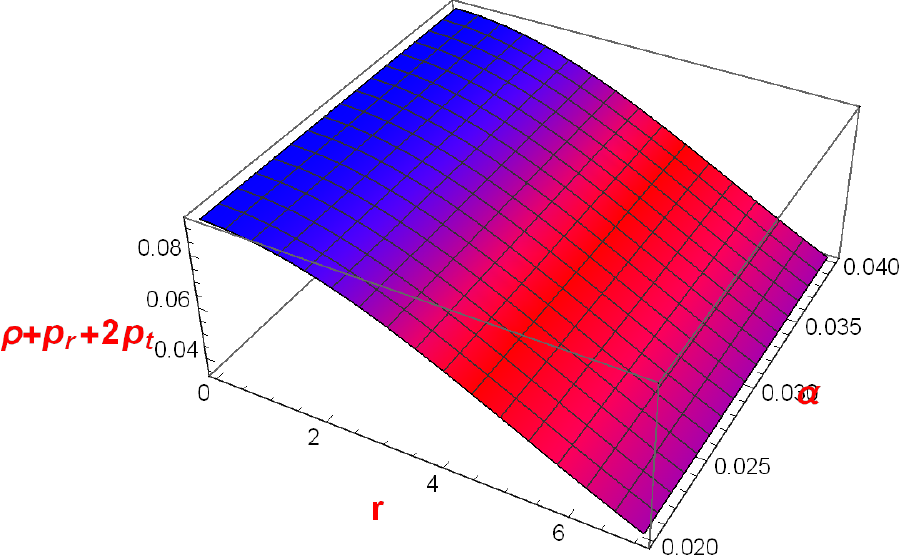, width=.32\linewidth,
height=2.5in}\epsfig{file=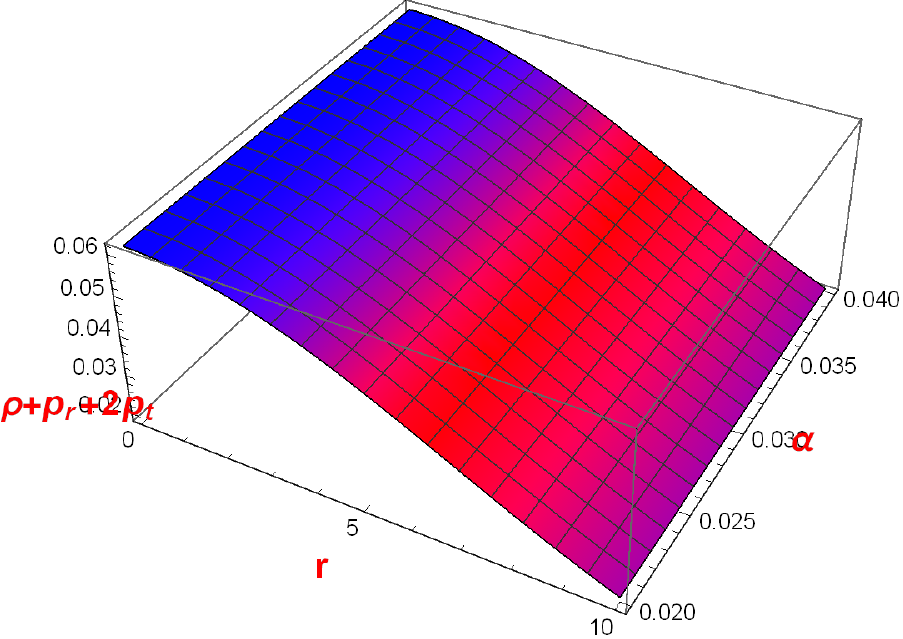, width=.32\linewidth,
height=2.5in}\caption{\label{Fig.19} shows the behavior of $\rho+p_{r}+2p_t$.}
\end{figure}

Energy conditions are important tools for the non-vacuum solution as well as in relativistic cosmology. In modified theories of gravity, commonly, four types of energy conditions are considered. These conditions can be written as
\begin{itemize}
\item Null energy condition (NEC): $\rho+p_r \geq 0,\,\,\ \rho+p_t \geq 0$.
\item Weak energy condition (WEC): $\rho\geq 0$ and $\rho+p_r \geq 0,\,\,\ \rho+p_t \geq 0$.
\item Dominant energy condition (DEC): $\rho> |p_r|,\,\,\ \rho>|p_t|$.
\item Strong energy condition (SEC): $\rho+p_r+2 p_t \geq 0$.
\end{itemize}
Moreover, the dominant energy condition suggests that the speed of light should be greater than equal to the speed of energy. To ensure the DEC, we have shown the profiles of $\rho-p_r \geq 0,\,\,\ \rho-p_t \geq 0$ in Figure \ref{Fig.17} and \ref{Fig.18}, respectively. It is observed that DEC is validated. Furthermore, NEC, WEC, and SEC are satisfied, and their profiles are presented in Figures \ref{Fig.15}, \ref{Fig.16}, and \ref{Fig.19}.

\subsection{Mass function, compactness function, and redshift function}\label{sec4g}
In the compact stellar object, the mass function $m(r)$, compactness function $u(r)$, and redshift function $Z_s$ play an important role in understanding its physical properties. The mass function for the stellar object can be written as
\begin{equation}
m(t)=\int_0^r (4\pi r^2 \rho) dr.
\end{equation}

The compactness function can be defined as
\begin{equation}
u(r)=\frac{2 m(r)}{r}.
\end{equation}

The gravitational redshift function for the stellar object can be written as

\begin{equation}
Z_s=\frac{1}{\sqrt{1-2u(r)}}\left(1-\sqrt{1-2 u(r)}\right)
\end{equation}

\begin{figure}[H]
\centering \epsfig{file=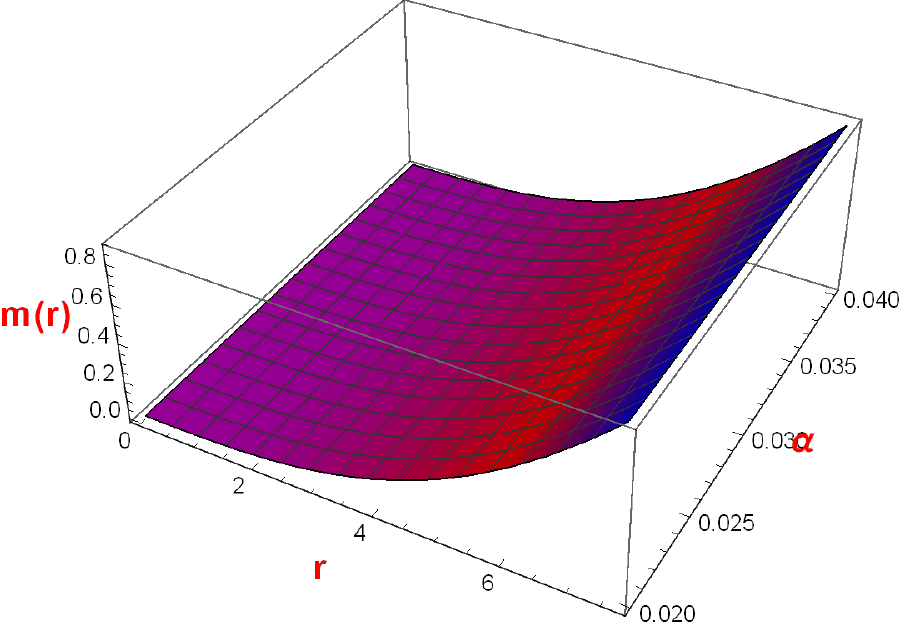, width=.32\linewidth,
height=2.5in}\epsfig{file=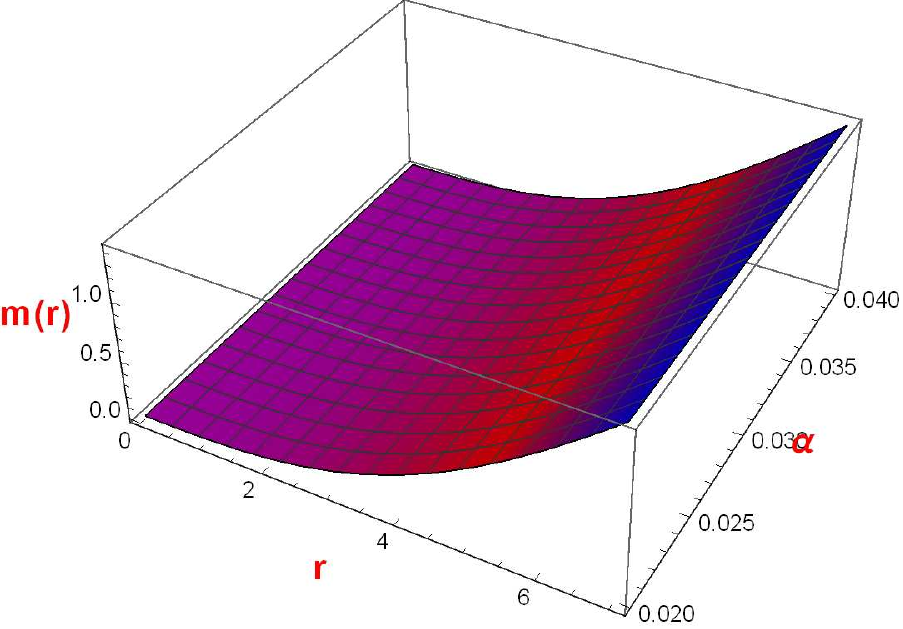, width=.32\linewidth,
height=2.5in}\epsfig{file=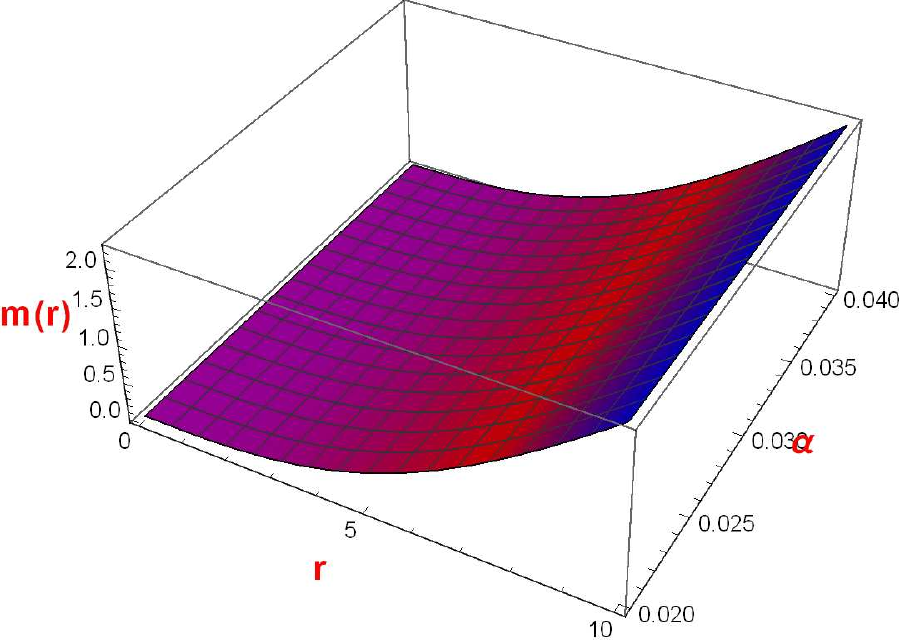, width=.32\linewidth,
height=2.5in}\caption{\label{Fig.21} shows the behavior of $m(r)$.}
\end{figure}

\begin{figure}[H]
\centering \epsfig{file=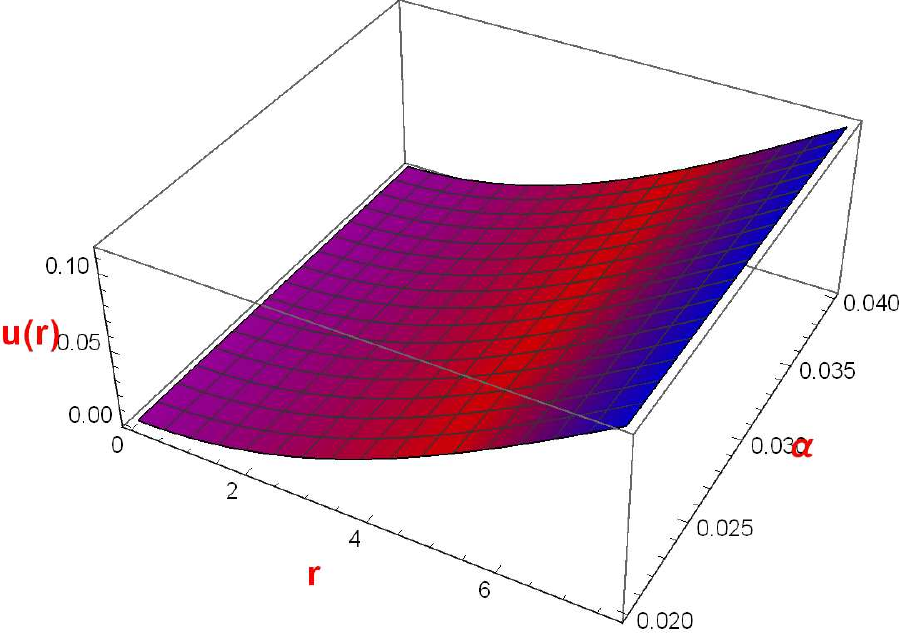, width=.32\linewidth,
height=2.5in}\epsfig{file=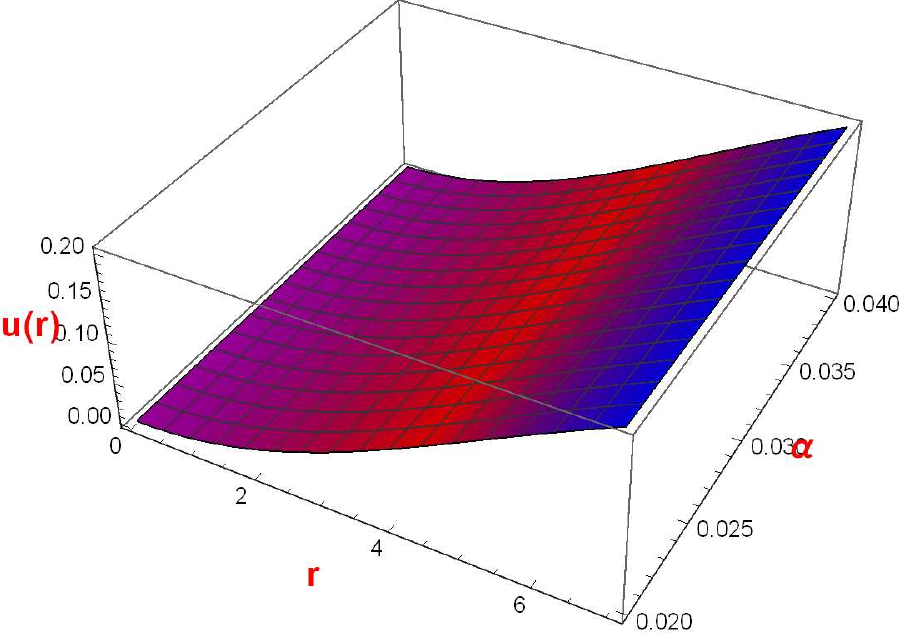, width=.32\linewidth,
height=2.5in}\epsfig{file=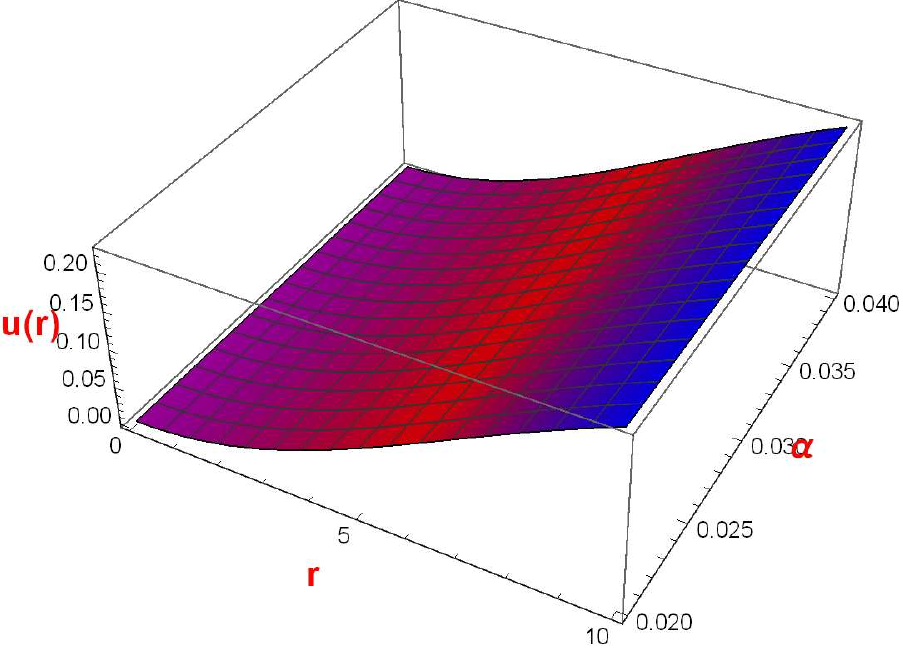, width=.32\linewidth,
height=2.5in}\caption{\label{Fig.22} shows the behavior of $u(r)$.}
\end{figure}

\begin{figure}[H]
\centering \epsfig{file=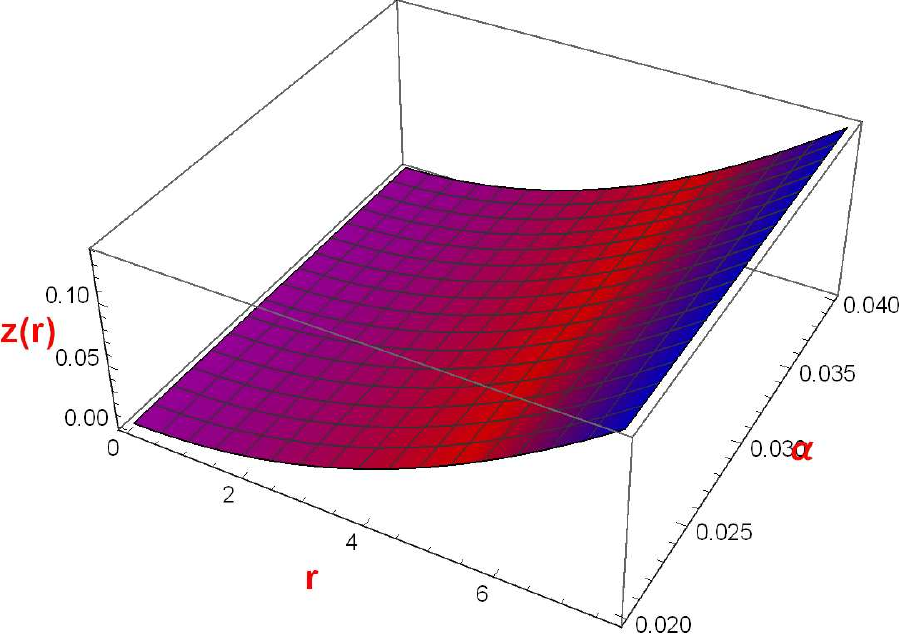, width=.32\linewidth,
height=2.5in}\epsfig{file=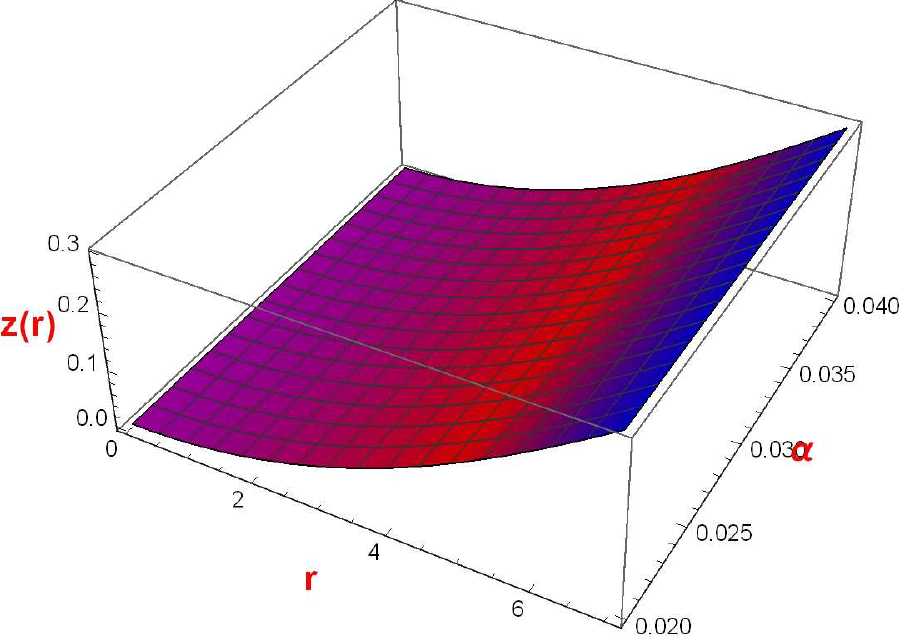, width=.32\linewidth,
height=2.5in}\epsfig{file=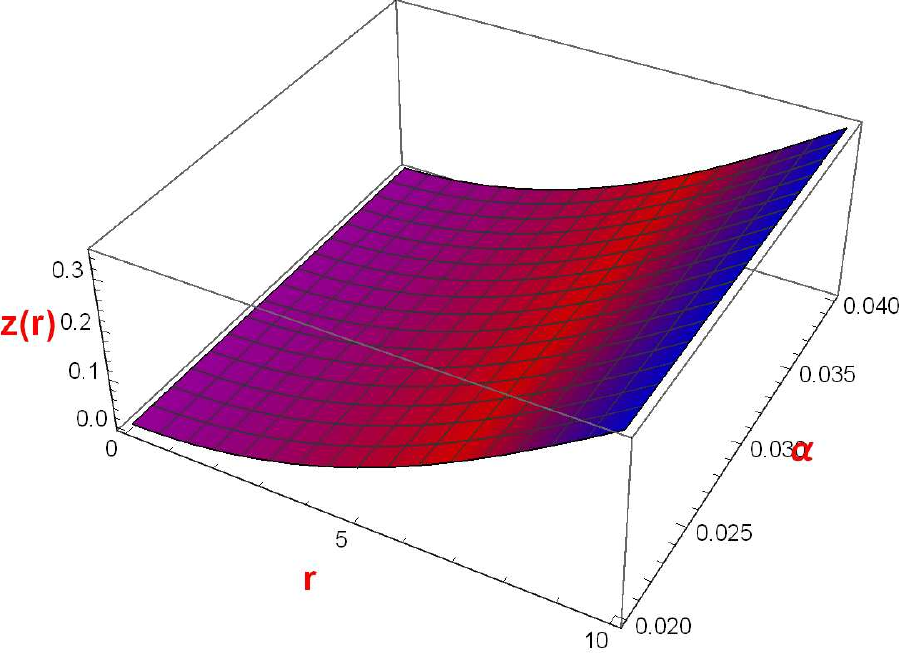, width=.32\linewidth,
height=2.5in}\caption{\label{Fig.23} shows the behavior of $z(r)$.}
\end{figure}

Figure \ref{Fig.21} illustrates the behavior of mass function $m(r)$, from which it is seen that the mass function increases monotonically from the center to the surface of compact star for three models. The compactness function's evolution is shown in Figure \ref{Fig.22}. We observed that $u(r)\leq 0.30$, which satisfied the Buchdahl condition \cite{Buchdhal/1959} for the current study. The gravitational redshift function $Z_s$ is depicted in Figure \ref{Fig.23} for three models. The values of $Z_s$ are less than 0.5 i.e., $Z_s <0.5$ for all considered models. These results are compatible with the results proposed in literature such as $Z_s\leq 5$ by Bohmer and Harko \cite{Bohmer/2006}, $Z_s\leq 5.211$ by Ivanov \cite{Ivanov/2002}, $Z_s\leq 2.400$  Mustafa, Shamir, and Tie-Cheng \cite{Mustafa/2020a} for anisotropy fluid sphere. All of the above results suggest that our compact star models are validated in the framework of modified $f(Q)$ gravity.

\section{Conclusion}\label{sec5}

In this manuscript, we studied the compact stellar objects in well-known modified $f(Q)$ gravity with the presence of a quintessence field. We focused on exploring three compact star models, namely, HerX-1, SAXJ1808.4-3658, and 4U1820-30, with the Schwarzchild spacetime and anisotropic source of fluid. To do this, we have considered an exponential form of $f(Q)$ as $f(Q)=Q-\alpha \gamma \left(1-\frac{1}{e^{\frac{Q}{\gamma}}}\right)$ and formulated the field equations in $f(Q)$ gravity. To our best of knowledge, this is the first attempt to study compact stars in the background of $f(Q)$ gravity. Furthermore, the matching condition is used to find the different values of parameters, i.e., the Schwarzchild metric as an exterior solution with the interior spacetime metric at the boundary. Besides this, the model parameter $\alpha$ act an important role in our analysis as the rest of our study depends on it, and we choose $\alpha$ in $[0.020, 0.040]$ to proceed our study. It is worthy to note here that we calculated the values of parameters by imposing the matching condition. In our further study, we have analyzed the profiles of energy-momentum components i.e., $\rho, \,\ p_r, \,\ p_t, \,\ \rho_q$, equation of state parameters ($\omega_r$ and $\omega_t$), gradients ( $\rho'(r),\,\ p_r'(r),\,\ p_t'(r)$), anisotropy function $\Delta=p_t-p_r$, causality and Abreu stability analysis, energy conditions, and mass function, compactness function, and redshift function. Some of the important results of this manuscripts are discussed in the following:
\begin{itemize}

\item The graphical representation of energy density $\rho$ is depicted in Figure \ref{Fig.1} for three compact star models in  $f(Q)$ gravity. It is observed that $\rho$ is positive throughout the configuration of all parameters, whereas the quintessence energy density is negative for the same configuration (see Figure \ref{Fig.4}). The radial and tangential pressures are positive for $r<R$, and their profiles are shown in Figures \ref{Fig.2} and \ref{Fig.3}.

\item The Equation of state parameter for radial and tangential pressure should be lied in between $0$ and $1$ to present a stable stellar object. From Figures \ref{Fig.5} and \ref{Fig.6}, we have seen that the EoS satisfied the setup for the stable stellar model.

\item The gradient of energy density, radial, and tangential pressures are also important for our analysis. The gradient must satisfy the following conditions\\
for $0<r\leq R$, we must have
$\frac{d\rho}{dr}<0,\,\,\ \frac{dp_r}{dr}<0,\,\,\ \frac{dp_t}{dr}<0$ and at the origin
$\frac{d\rho}{dr}\bigg|_{r=0}=0,\,\,\ \frac{dp_r}{dr}\bigg|_{r=0}=0,\,\,\ \frac{dp_t}{dr}\bigg|_{r=0}=0.$\\
The profiles of all of the above conditions are presented in Figures \ref{Fig.7}, \ref{Fig.8}, and \ref{Fig.9}.

\item The behavior of the anisotropic function is illustrated in Figure \ref{Fig.10} for three stellar models. The positive behavior of anisotropic function suggests that the anisotropic force is repulsive because $p_t>p_r$.

\item From Figures \ref{Fig.11} and \ref{Fig.12}, we depicted the radial and tangential velocities for three models. It is seen that the velocities are satisfied the condition $0\leq v_r \,\ \& \,\ v_t \leq 1$. Also, the Abrea condition is presented in Figures \ref{Fig.13} and \ref{Fig.14}.

\item Moreover, all the energy conditions such as NEC, WEC, DEC, and SEC are satisfied for three stellar models. Furthermore, their behavior is shown in Figures \ref{Fig.1}, \ref{Fig.15}, \ref{Fig.16}, \ref{Fig.17}, \ref{Fig.18}, and \ref{Fig.19}.

\item Furthermore, the graphical representation of mass function $m(r)$, compactness function $u(r)$, and redshift function $Z_s$ have been shown in Figures \ref{Fig.21}, \ref{Fig.22}, and \ref{Fig.23}, respectively. In this study, we observed that the mass function monotonically increases towards the surface of stellar structure. The compactness function also satisfied the Buchdahl condition i.e., $u(r)\leq 0.30$. In addition, we analyze the redshift function for $Z\leq 0.15$, which is a good agreement with the results presented by \cite{Buchdhal/1959,Bohmer/2006, Ivanov/2002, Mustafa/2020a}.
\end{itemize}

Finally, combining all the outcomes, it is concluded that the discussed stellar models are physically acceptable in the framework of modified $f(Q)$ gravity. It would be interesting to extend the analysis of compact stars in $f(Q)$ gravity in the near future.

\section*{Acknowledgments}

S.M. acknowledges Department of Science \& Technology (DST), Govt. of India, New Delhi, for awarding Senior Research Fellowship (File No. DST/INSPIRE Fellowship/2018/IF180676). Z.H. acknowledges Department of Science and Technology (DST), Government of India, New Delhi, for awarding a Junior Research Fellowship (File No. DST/INSPIRE Fellowship/2020/IF190911). We are very much grateful to the honorable referee and the editor for the illuminating suggestions that have significantly improved our work in terms of research quality and presentation.

\end{document}